\documentclass[prb,aps,twocolumn,showpacs,superscriptaddress]{revtex4-2}
\usepackage{amsmath,graphicx,amssymb,braket,enumitem,mathtools,dsfont}
\usepackage[colorlinks=true,citecolor=blue,linkcolor=red]{hyperref}
\usepackage{titlesec}
\usepackage{booktabs}
\usepackage{siunitx}
\usepackage{bbold}
\usepackage{relsize}

\usepackage{tikz}
\usetikzlibrary{calc}
\usetikzlibrary{positioning}
\usetikzlibrary{decorations.pathreplacing}

\usepackage[T1]{fontenc}

\usepackage{gensymb}

\usepackage{subcaption,cleveref}

\newcommand{\overbar}[1]{\mkern 1.5mu\overline{\mkern-1.5mu#1\mkern-1.5mu}\mkern 1.5mu}

\titlespacing{\subsection}
    {0pt}{9pt}{6pt}

\graphicspath{{Figures}}

\begin{document}

\title{3D tomography of exchange phase in a Si/SiGe quantum dot device}
\author{Dylan Albrecht}
\affiliation{Sandia National Laboratories, Albuquerque NM, USA}
\author{Sarah Thompson}
\affiliation{Sandia National Laboratories, Albuquerque NM, USA}
\author{N. Tobias Jacobson}
\affiliation{Sandia National Laboratories, Albuquerque NM, USA}
\author{Ryan Jock}
\affiliation{Sandia National Laboratories, Albuquerque NM, USA}

\date{\today}

\begin{abstract}
The exchange interaction is a foundational building block for the operation of spin-based quantum processors. Extracting the exchange interaction coefficient $J(\mathbf{V})$, as a function of gate electrode voltages, is important for understanding disorder, faithfully simulating device performance, and operating spin qubits with high fidelity. Typical coherent measurements of exchange in spin qubit devices yield a modulated cosine of an accumulated phase, which in turn is the time integral of exchange. As such, extracting $J(\mathbf{V})$ from experimental data is difficult due to the ambiguity of inverting a cosine, the sensitivity to noise when unwrapping phase, as well as the problem of inverting the integral. As a step toward obtaining $J(\mathbf{V})$, we tackle the first two challenges to reveal the accumulated phase, $\phi(\mathbf{V})$.
We incorporate techniques from a wide range of fields to robustly extract and model a 3D phase volume for spin qubit devices from a sequence of 2D measurements. In particular, we present a measurement technique to obtain the wrapped phase, as done in phase-shifting digital holography, and utilize the max-flow/min-cut phase unwrapping method (PUMA) to unwrap the phase in 3D voltage space.
We show that this method is robust to the minimal amount of observed drift in the device, which we confirm by increasing scan resolution. Upon building a model of the extracted phase information, we optimize over the model to locate a minimal-gradient $\pi$ exchange pulse point in voltage space.
We propose automating the presented procedure to model many exchange axes in other devices and tunings.
Our measurement protocol may provide detailed information useful for understanding the origins of device variability governing device yield, enable calibrating device models to specific devices during operation for more sophisticated error attribution, and enable a systematic optimization of qubit control.  We anticipate that the methods presented here may be applicable to other qubit platforms.
\end{abstract}

\maketitle
\clearpage

\section{Introduction}
\noindent
Spin qubits based on gate-defined quantum dots (QDs) in silicon are attractive for implementing large-scale quantum information processors due to their compatibility with industrial silicon foundry processing technology\cite{Koch2025, Steinacker2025, Neyens2024, Engel2001, Ladd2010}. 
In these platforms, information is encoded using the spins of electrons confined in individual QDs. 
While silicon spin qubits have successfully demonstrated multi-qubit gate operation~\cite{deFuentes2026, Madzik2025operating, Zhang2025, Fedele2021} utilizing a range of qubit encodings, continued progress toward scalable, fault-tolerant quantum computing likely requires deep understanding of the physical error mechanisms governing qubit performance and how these mechanisms relate to device fabrication details and qubit control. 
Operation of many silicon spin qubits may require that control of the exchange interaction between electrons on neighboring QDs be realized with high precision and consistency across many qubit sites.
These interactions are sensitive to nanometer-scale variations in a given device structure, and are therefore unique to each interacting pair of QDs in a device~\cite{Ha2022}.
This sensitivity makes exchange a useful probe of local disorder~\cite{Tariq2022, Buterakos2021}, while presenting difficulties for device calibration and performance.

Understanding and controlling exchange interactions in silicon spin qubits requires mapping the parameters of an applied voltage pulse at the gate electrode onto the induced electron spin-spin exchange interaction. This coupling may be inferred from the associated integrated phase. 
The typical observable in experiments designed to calibrate or study exchange interactions is a trigonometric function of a time- and voltage-dependent accumulated phase, which makes direct analysis of the voltage dependence non-trivial.
A typical approach for mapping voltage onto phase using this measurement output, which has proven sufficient for calibrating control pulses, involves either directly fitting the output signal to an analytical expression~\cite{Madzik2025operating}, or extracting the voltage coordinates of inflection points in the output signal corresponding to integer accumulations of $\pi$ rotations, then fitting an analytical expression to those coordinates and interpolating to estimate the voltage corresponding to any other phase~\cite{Andrews2019}. 
The former is a nontrivial, nonlinear optimization fitting procedure and can be thought of as a 1D phase unwrapping followed by a fitting of the result to a high-order polynomial, which is a simpler convex optimization problem. The difficulties of nonlinear fitting become more apparent when the analysis is extended to a higher-dimensional voltage space. 
The calibration analysis addresses only a cross-section of the full voltage space during an exchange pulse, which typically has three dimensions corresponding to the three gate electrodes used to define a pair of QDs and the potential barrier between them.
These three gates must be pulsed simultaneously as the exchange interaction is modulated so that the electrostatic potential difference (detuning) between the QDs is maintained (or controlled), and the capacitive effect of barrier gate pulses on the QDs is compensated. Voltage noise and pulse distortion effects, such as skew, can pull the pulse off of this line of (symmetric) operation. This is important for simulating the effects of noise on the gate electrodes, which requires a reasonably accurate model for $J(\mathbf{V})$. 
Therefore, for important practical and analytical objectives, a higher-dimensional, direct phase unwrapping approach is desirable. 
Phase unwrapping in the full, 3D voltage space may guide tailored qubit control voltage trajectories that optimize exchange phase coherence in the presence of device-specific non-idealities. 
An accurate inference of voltage-dependent phase may furthermore facilitate comparison with simulated data from microscopic physical models so that the local electronic structure giving rise to such non-idealities may be understood.
Finally, we note that direct 3D phase unwrapping can facilitate calibrations for the impulse response of the control signal path, which is difficult to measure given practical constraints imposed by millikelvin operating temperatures within a dilution refrigerator~\cite{Barnes2020, Ni2025, Rol2020}.

Similar phase unwrapping problems appear and are analyzed in many different scientific areas -- interferometric synthetic aperture radar (SAR/inSAR)~\cite{Yu2019}, magnetic resonance imaging (MRI)~\cite{Jenkinson2003,Jenkinson2012}, optical interferometry~\cite{Aiello2007,Zuo2022}, and fringe projection profilometry (FPP)~\cite{Su2001,Hu2020,Feng2021}. In navigating the literature, it is important to distinguish between cases where the wrapped phase is readily available due to the measurement technique, cases where the phase is an argument of a trigonometric function, and cases where the analysis relies on an input carrier frequency or other additional data, such as a coherence map~\cite{Yu2019}. In most applications, including ours, it is necessary to obtain the true phase, requiring an appropriate unwrapping technique. Because of this, and the fact that phase unwrapping is a nontrivial problem, especially when considering robustness to a variety of different types of noise, researchers in disparate fields have generated a large body of work on phase unwrapping algorithms over the past 30+ years~\cite{Judge1994,Wang2022}. Here, we leverage this body of work to present a reliable method for obtaining and modeling exchange phase of an encoded silicon spin qubit in 3D voltage space. In particular, we use the well-known max-flow/min-cut phase unwrapping method (PUMA)~\cite{Bioucas-Dias2007}, which requires the $(-\pi,\pi)$ wrapped phase directly as input, to perform phase unwrapping of a collection of 2D raster scans.
We present our procedure in two parts, the first of which deals with the experimental protocol to extract a $(-\pi,\pi)$ wrapped phase from typical measurement outputs. The second part deals with the application of the phase unwrapping algorithm and building a phase volume model.

\section{Qubit device and data collection}

\begin{figure*}[!ht]
\centering
\begin{tabular}{cc}
\subfloat[\label{fig:DeviceIntro:a}]{\includegraphics[width=50mm]{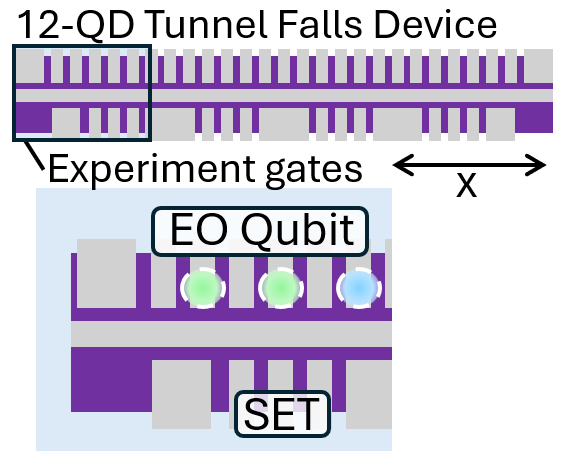}} &
\subfloat[\label{fig:DeviceIntro:b}]{
\includegraphics[width=70mm]{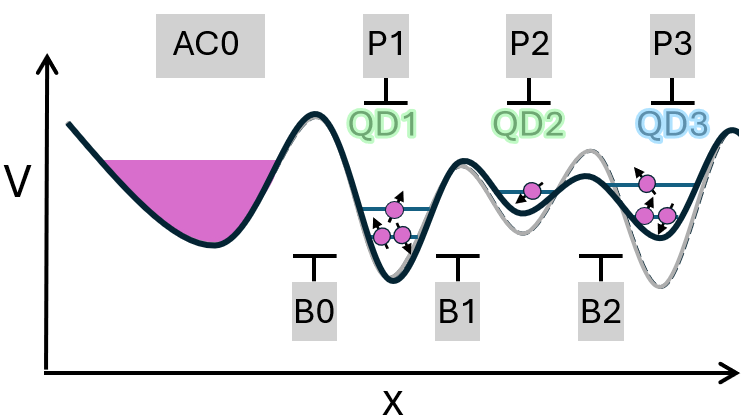}}
\\
\subfloat[
\label{fig:DeviceIntro:c}]{\includegraphics[width=27mm]{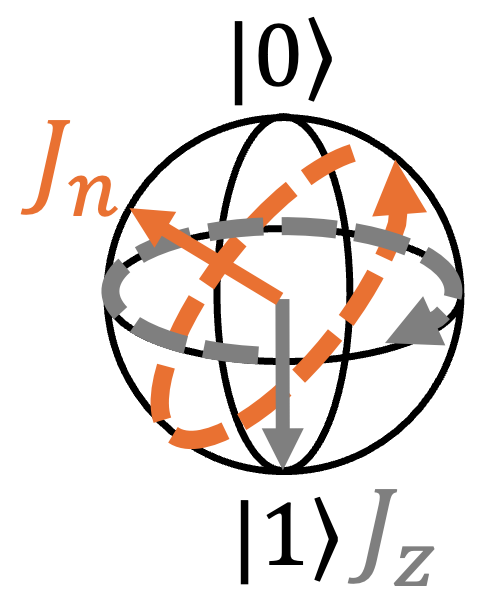}} &
\subfloat[\label{fig:DeviceIntro:d}]{
\includegraphics[width=90mm]{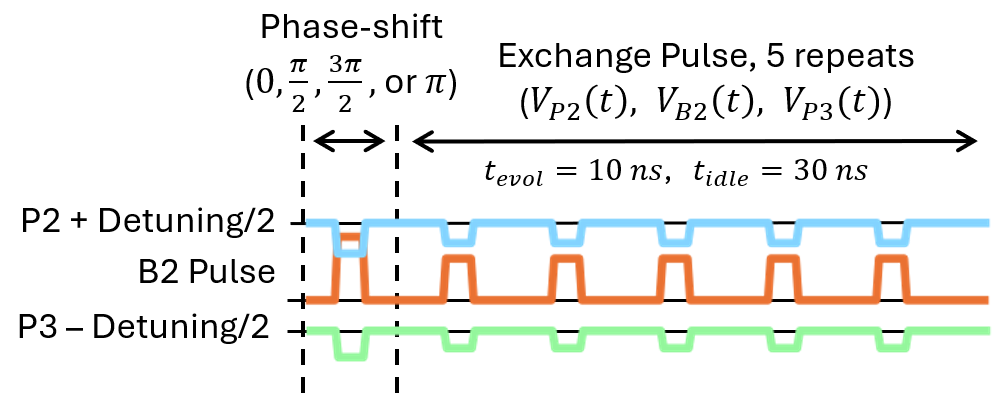}}
\end{tabular} 
\caption{
\textbf{Device operation.} (a) Schematic illustration of a 12-QD Intel Si/SiGe Tunnel Falls device with magnified view of device region used for this experiment. QDs are formed under plunger gates (P1, P2, P3) on the top half of the device, as indicated by green (addressed during state preparation and measurement) and blue (gauge spin) circles, and are separated by barrier gates (B0, B1, B2). (b) Schematic electrostatic potential diagram for the experimental section of the device under idle conditions (grey) and during an exchange pulse (black). (c) Qubit bloch sphere, where $J_n$ is the exchange interaction between QD2 and QD3, and $J_z$ is the exchange interaction between QD1 and QD2. (d) Example pulse control sequence during a phase-shifted fingerprint measurement.
}
\label{fig:DeviceIntro}
\end{figure*}

\noindent
In this work, we utilize an industrially manufactured, isotopically enriched (800 ppm \textsuperscript{29}Si) silicon-based device, depicted in \ref{fig:DeviceIntro:a}, to measure phase between electron spins as a function of voltage pulse-driven exchange interactions. The device was fabricated at Intel Corp.'s quantum device foundry (similar to devices used in \cite{Neyens2024, marcks2025valley,Mądzik2025}). We operate the system in a three-electron-spin exchange-only (EO) qubit encoding \cite{PhysRevLett.116.110402}, illustrated in \ref{fig:DeviceIntro:b}. 
We initialize the qubit in the logical $|0\rangle$ state by preparing a spin singlet on QD1 and QD2, with a total spin of $S=1/2$ and a (3, 1, 3) charge configuration across all three QDs~\cite{Reed2016}. 
Coherent qubit control is achieved by applying base-band voltage pulses to the QD plunger and barrier gates to turn on and off the Heisenberg exchange interaction, $J$, between neighboring QDs.
The Heisenberg interaction Hamiltonian for the system of single electrons occupying each QD is
\begin{equation}
H = J_{12}\left(\mathbf{V}(t)\right) \; \mathbf{S}_{1} \cdot \mathbf{S}_{2} + J_{23} \left(\mathbf{V}(t)\right) \; \mathbf{S}_{2} \cdot \mathbf{S}_{3}
\end{equation}
where $J_{z} \equiv J_{12}$ and $J_{n}\equiv J_{23}$ are the exchange interactions between QDs 1-2 and 2-3, for the $z$ and $n$ axes, respectively, and $\mathbf{S}_{i}$ are spin operators corresponding to the $i^{\rm th}$ QD. 
Following manipulation by the sequence of exchange pulses, the qubit state is measured through Pauli spin blockade (PSB), which allows spin-dependent electron transfer between QD1 and QD2 to be sensed by capacitive coupling to a single electron transistor (SET), corresponding to a projection onto the z-axis in \ref{fig:DeviceIntro:c}~\cite{PRXQuantum.4.010329}. 
The measurement outcome is reported as the probability of singlet return, \emph{i.e.}~the probability of measuring that the qubit has returned to the originally prepared logical $|0\rangle$ state following coherent manipulation.
In this work we focus on pulses to the P2, B2, and P3 gates, to invoke the exchange interaction between the spin on QD2 and the spin on QD3 (typically referred to as the gauge spin)~\cite{PhysRevLett.116.110402}. 
This interaction introduces a rotational phase $\phi$ around the $J_{n}$-axis of the Bloch sphere (Figures \ref{fig:DeviceIntro:b}-\ref{fig:DeviceIntro:d}). The singlet return probability yields a measurement of $\cos(\phi)$, and is a function of the time-integrated exchange interaction $J_n$ between spins on QD2 and QD3 during manipulation. For interpreting the data used in the following analysis, it is important to note that the singlet return probability ideally oscillates between 1 and 0.25 (rather than 1 and 0) due to the inherent non-orthogonality between the prepared state and the coherent control axis in \ref{fig:DeviceIntro:c} {\cite{DiVincenzo2000, Andrews2019}}. 

\begin{figure*}[!ht]
\centering
\begin{tabular}{cc}
\subfloat[\label{fig:experimental_scans:a}]{\includegraphics[width=61mm]{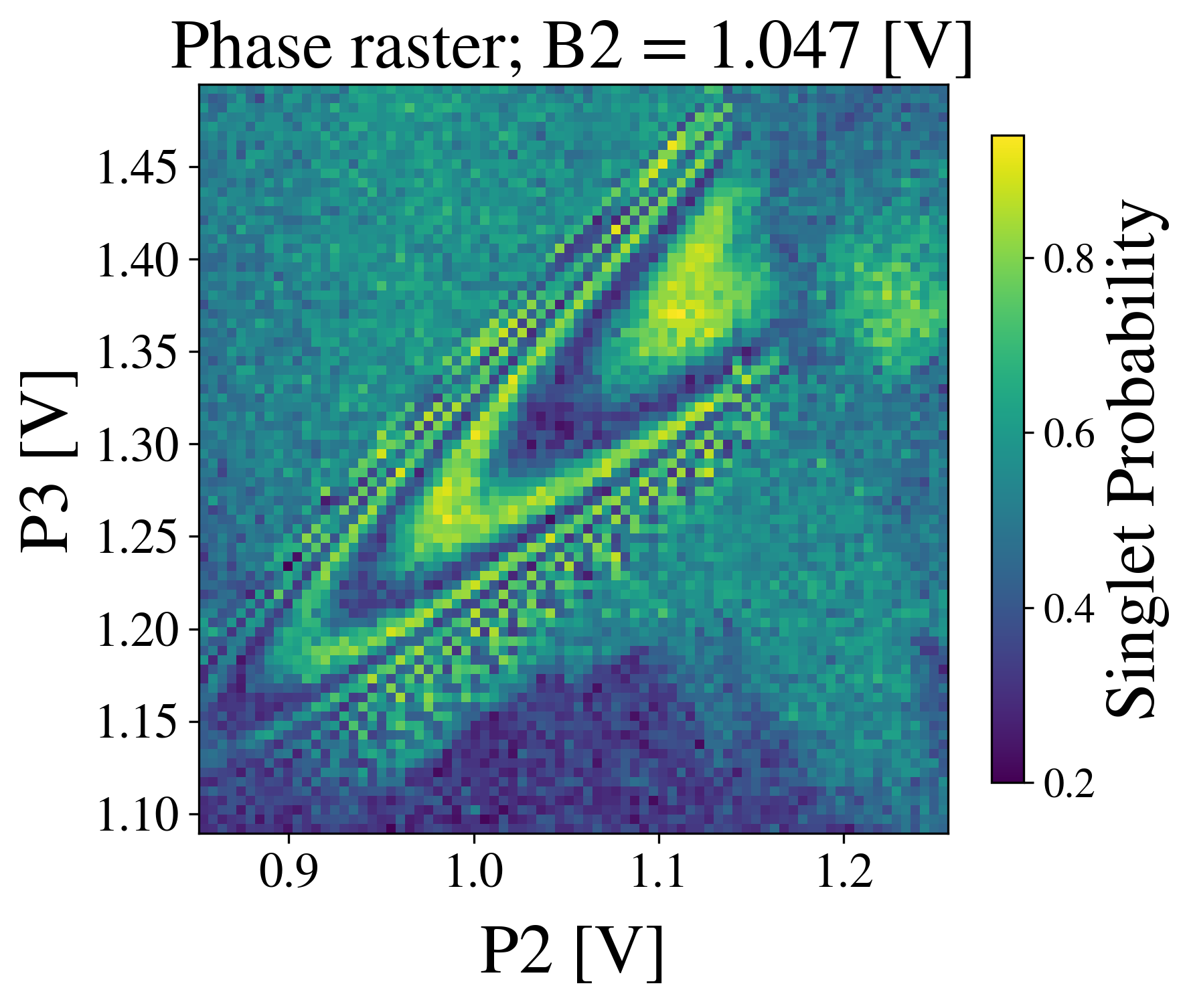}} &
\subfloat[\label{fig:experimental_scans:b}]{
\includegraphics[width=70mm]{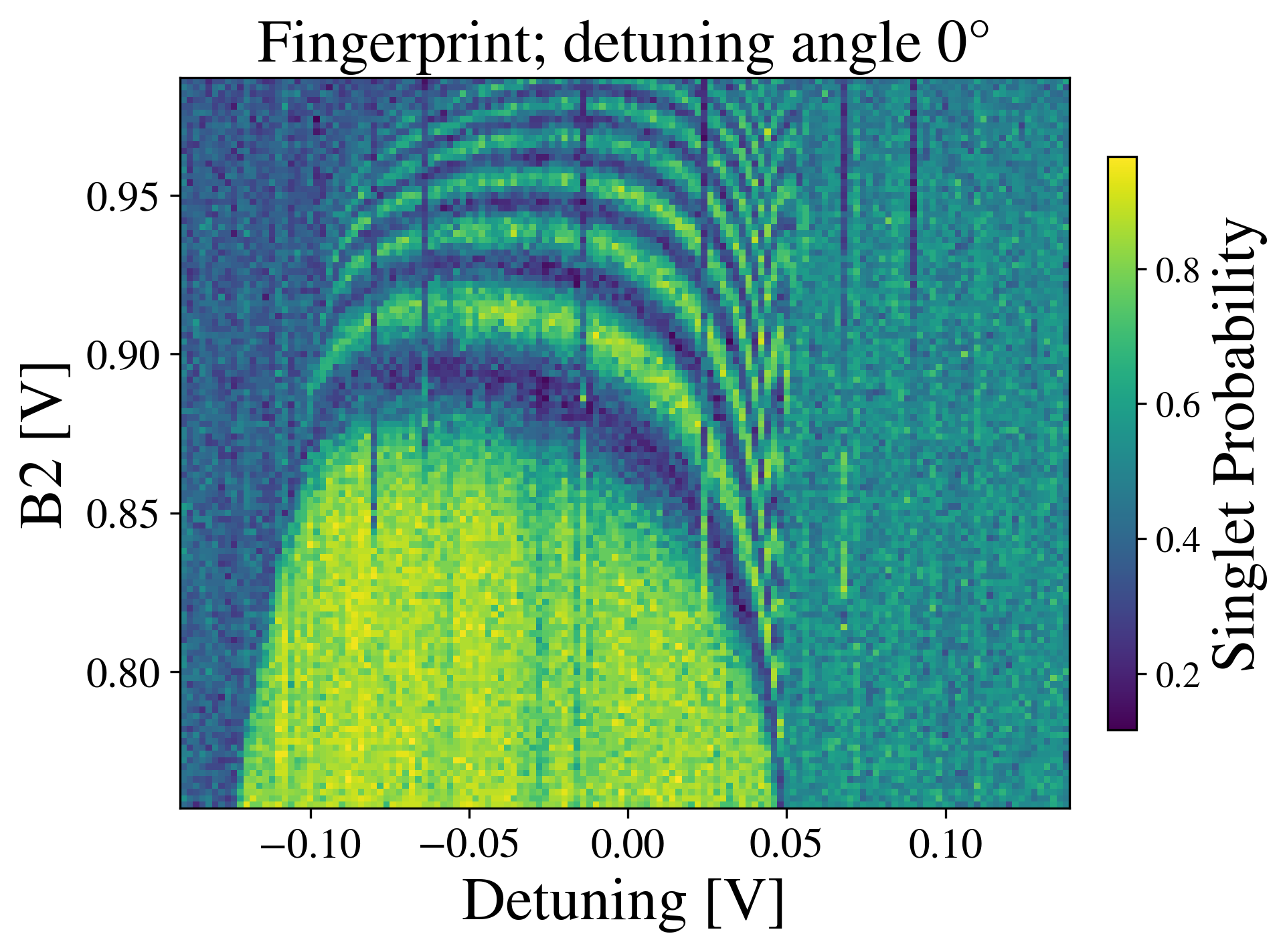}}
\\
\subfloat[
\label{fig:experimental_scans:c}]{\includegraphics[width=75mm]{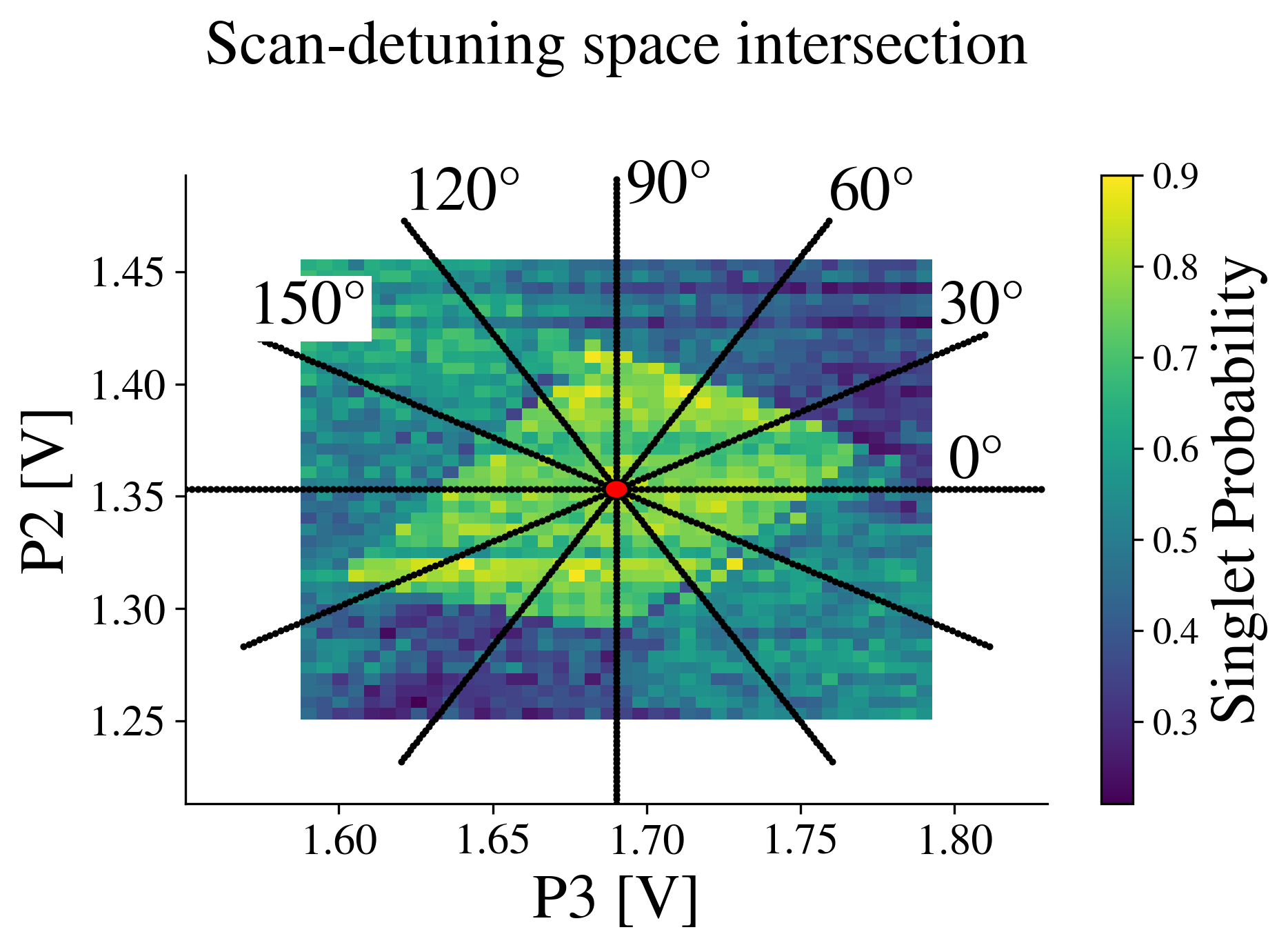}} &
\subfloat[\label{fig:experimental_scans:d}]{
\includegraphics[width=56mm]{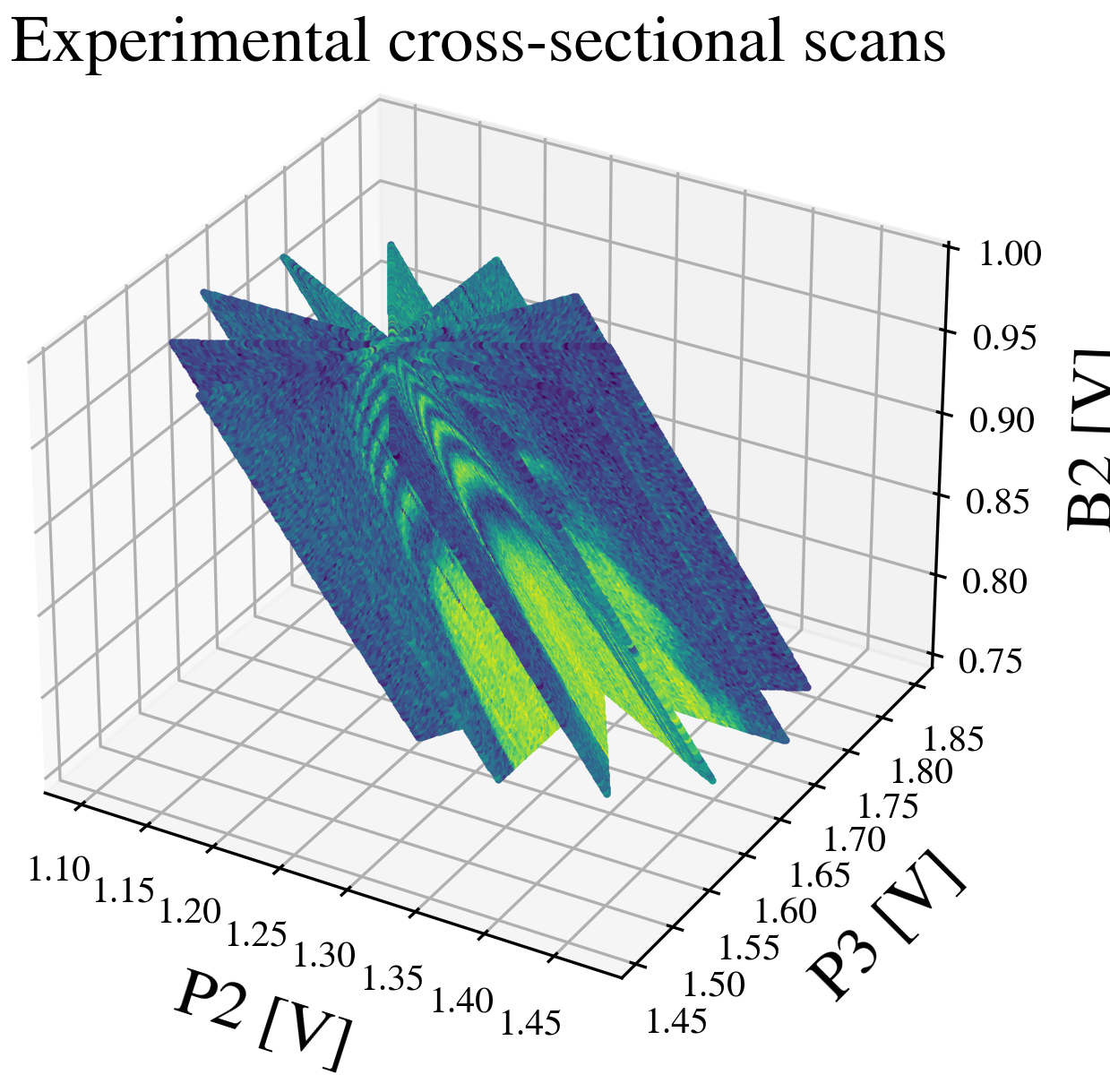}}
\end{tabular} 
\caption{
\textbf{Measurements.}
In figure (a) we have an example of a phase raster scan at fixed $V_{B2}$ greater than $V_{B2_{\textrm{idle}}}$.  In figure (b) we have an example of a fingerprint scan taken at $0\degree$ rotation in detuning space, c.f. (c).  In figure (c) we show the trajectories through detuning space used to collect fingerprints in (d). Each black line indicates the $V_{P2}$ and $V_{P3}$ coordinates along the detuning ($V_{P2}$-$V_{P3}$) axis of a fingerprint corresponding to that detuning angle. Trajectories are plotted over a phase raster measured with $V_{B2} = V_{B2_{\textrm{idle}}}$ and prolonged wait times to show the stable detuning limits for this charge state, for reference. The red circle indicates a cylindrical core of data removed when model fitting.  Figure (d) shows the fingerprint scans as cross-sectional scans in 3D voltage space, for $0\pi$ phase shift.
}
\label{fig:experimental_scans}
\end{figure*}

One approach to visualizing the effect of exchange-driven rotations on the singlet return probability is to sweep the pulsed-plunger voltages $V_{P2}$ and $V_{P3}$ independently at a fixed pulsed-barrier voltage, $V_{B2}$. This procedure results in a two dimensional plot of acquired qubit phase, or non-equilibrium charge cell phase raster scan~\cite{Lanza2022,Acuna2022}, shown in Figure~\ref{fig:experimental_scans:a}, which appears as a voltage-chirped fringe pattern due to functional form of the phase. 
Another approach is to ramp $V_{B2}$ while sweeping the detuning between the plunger voltages ($V_{P2}$ - $V_{P3}$), creating a so-called fingerprint plot~\cite{Reed2016}, with a fringe pattern similar to the phase raster, as shown in Figure~\ref{fig:experimental_scans:b}. 
Here, we collect data in a series of two-dimensional fingerprint scans, rotating successively by $30\degree$ about an experimentally determined central point in $V_{P2}$ and $V_{P3}$ detuning space, along a nominal compensation trajectory, as indicated in Figure~\ref{fig:experimental_scans:c}. 
The result is a series of cross-sectional images, analogous to a computed tomography (CT) scan, as shown in Figure~\ref{fig:experimental_scans:d}.
Whether a series of phase rasters or a series of fingerprints is measured, the same phase volume information is obtained, represented (or sliced) in different ways. 
We prefer to work with fingerprint scans since they more naturally include the idle point (zero phase), which facilitates phase contour identification; in contrast, in phase rasters, like the one shown in Figure \ref{fig:experimental_scans:a}, it is unclear which fringe corresponds to which phase, and for a 3D volumetric reconstruction of phase this information would need to be tracked and reconciled across scans.

In order to invert the trigonometric expression of the singlet return probability and obtain the wrapped phase, it is helpful to have phase-shifted versions of the measurement. Here, we were originally inspired by the work of~\cite{Tounsi2019}, where one only needs a single fringe pattern -- they make use of the higher-order components of the Riesz transform to compute the phase-shifted ($0\pi$, $\pi/2$, $\pi$, $3\pi/2$) images. The Riesz transform is the higher-dimensional generalization of the Hilbert transform, or phase shifter. We found this method was not effective in our data analysis, possibly due to the amount and character of the noise in our experimental data, since the method works well on simulated fingerprints without noise or including small Gaussian noise contributions. Alternatively, one can calculate the hypercomplex monogenic signal, which is the generalization of the one-dimensional analytic signal to two dimensions, and attempt the inversion using only the first-order Riesz kernels~\cite{Felsberg2002,Sierra-Vazquez2010,Seelamantula2012,Fong2022}. Alternatively, again, the measurement technique in the area of digital holography uses phase-shifted interferometry to collect a series of phase-shifted measurements, say by ($0\pi$, $\pi/2$, $\pi$, $3\pi/2$), which are then combined in analysis~\cite{Yamaguchi1997,Zhou2009,Zhou2016}. This is the method we have found to be most successful for our data. There are other combinations of phase shifts possible, such as optical triangulation, that could be interesting to investigate as well~\cite{Fong2022}.

We therefore obtain, for each $V_{P2}$ and $V_{P3}$ detuning angle condition ($0\degree$, $30\degree$, $60\degree$, $90\degree$, $120\degree$, $150\degree$), a series of four phase-shifted measurements by applying the pulse sequence in \ref{fig:DeviceIntro:d}, in which a single pulse is used to impart a phase shift of $0\pi$, $\pi / 2$, $\pi$, or $3\pi/2$, followed by a typical fingerprint pulse sequence made up of 5 repeated pulses to that voltage point, to accentuate the acquired phase. The combination of detuning angle and phase shift conditions results in a total of $6 \times 4 = 24$ scans, each scan having $115\times140$ pixels, with a total data acquisition time of approximately 21 hours.

One consideration in choosing an experimental measurement protocol is sensitivity to drift in the device calibration, which can occur during and between phase-shifted scan variants, potentially resulting in erroneous wrapped phase extraction (this issue is avoided in the Riesz transform method, as emphasized in \cite{Tounsi2019}). Our chosen procedure for wrapped phase extraction and subsequent phase unwrapping, which we discuss next, appears to be robust to the minimal amount of observed drift in the device, which we measure as discussed in Section \ref{sec:supplement}. We check this by increasing our scan resolution to $300\times300$ pixels, taking roughly 3 hours for a single scan, or 12 hours for the phase-shifted set of four (phase raster) scans. This increased exposure to drift did not prevent extraction of the underlying phase information, as shown in Figure \ref{fig:hires_experimental_scans:b}. 

\section{Data analysis and modeling}
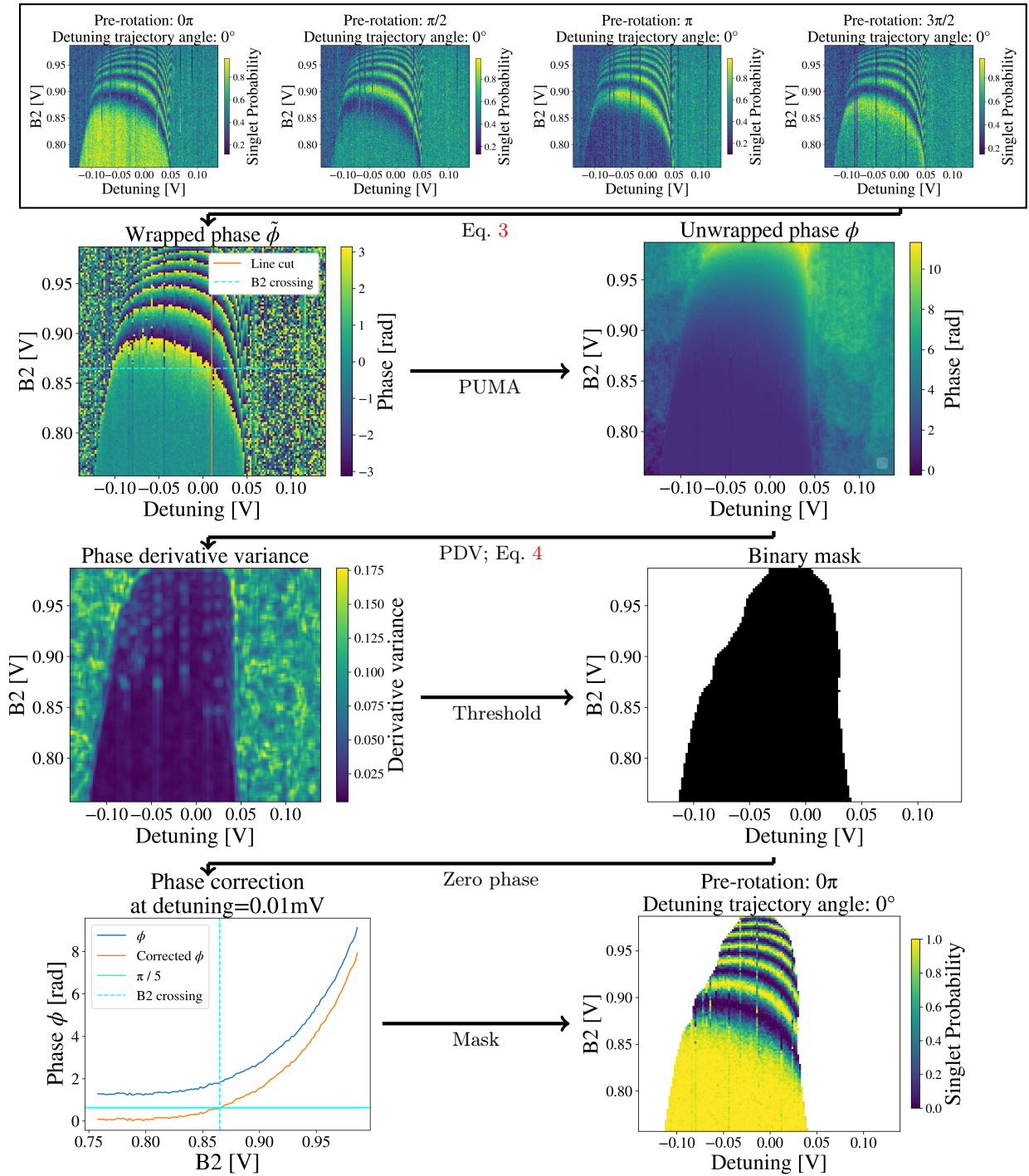
\begin{figure*}[!ht]
\centering
\begin{tikzpicture}[node distance=0cm]
\node (img1) {\includegraphics[width=3.8cm]{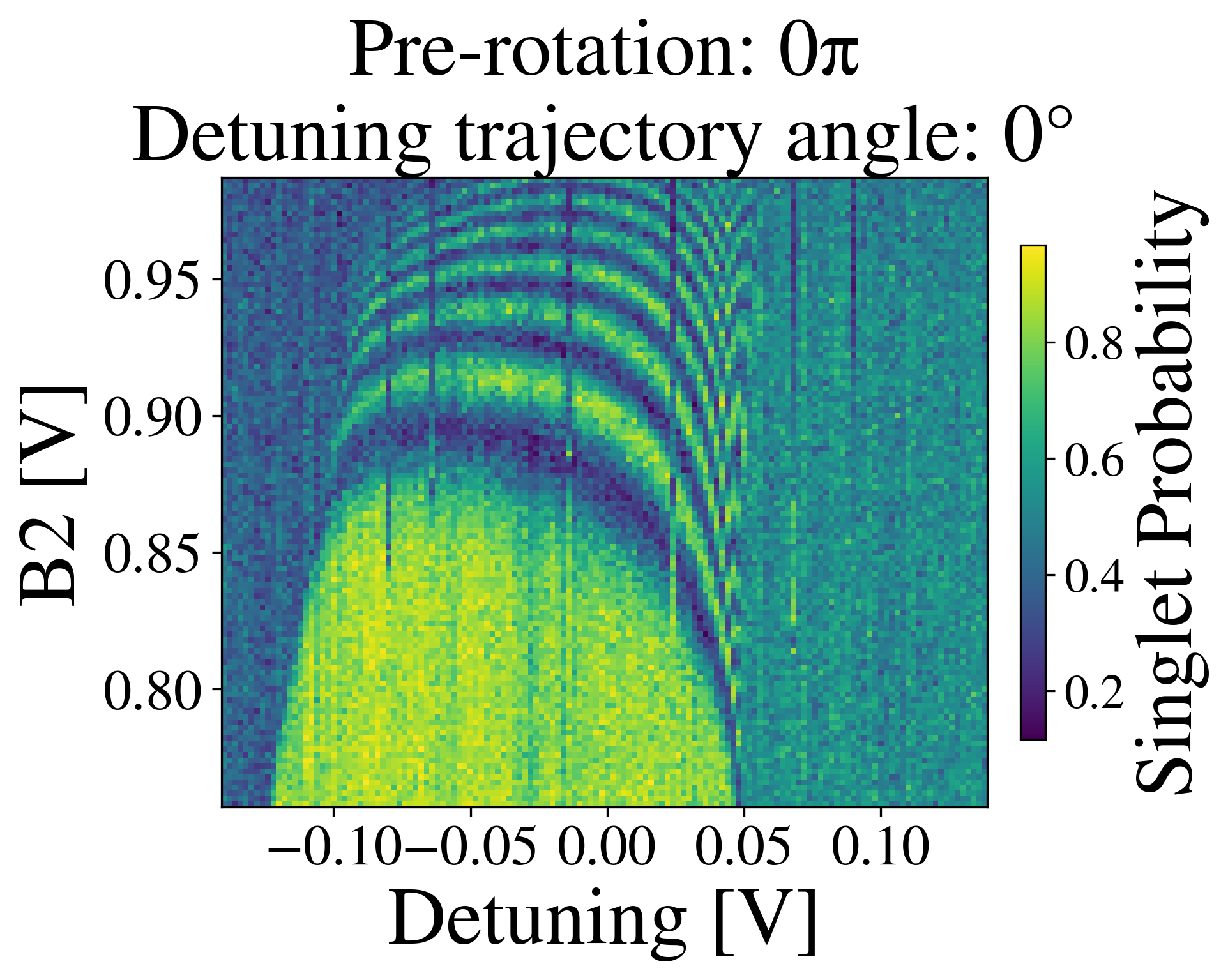}};
\node[right=of img1] (img2) {\includegraphics[width=3.8cm]{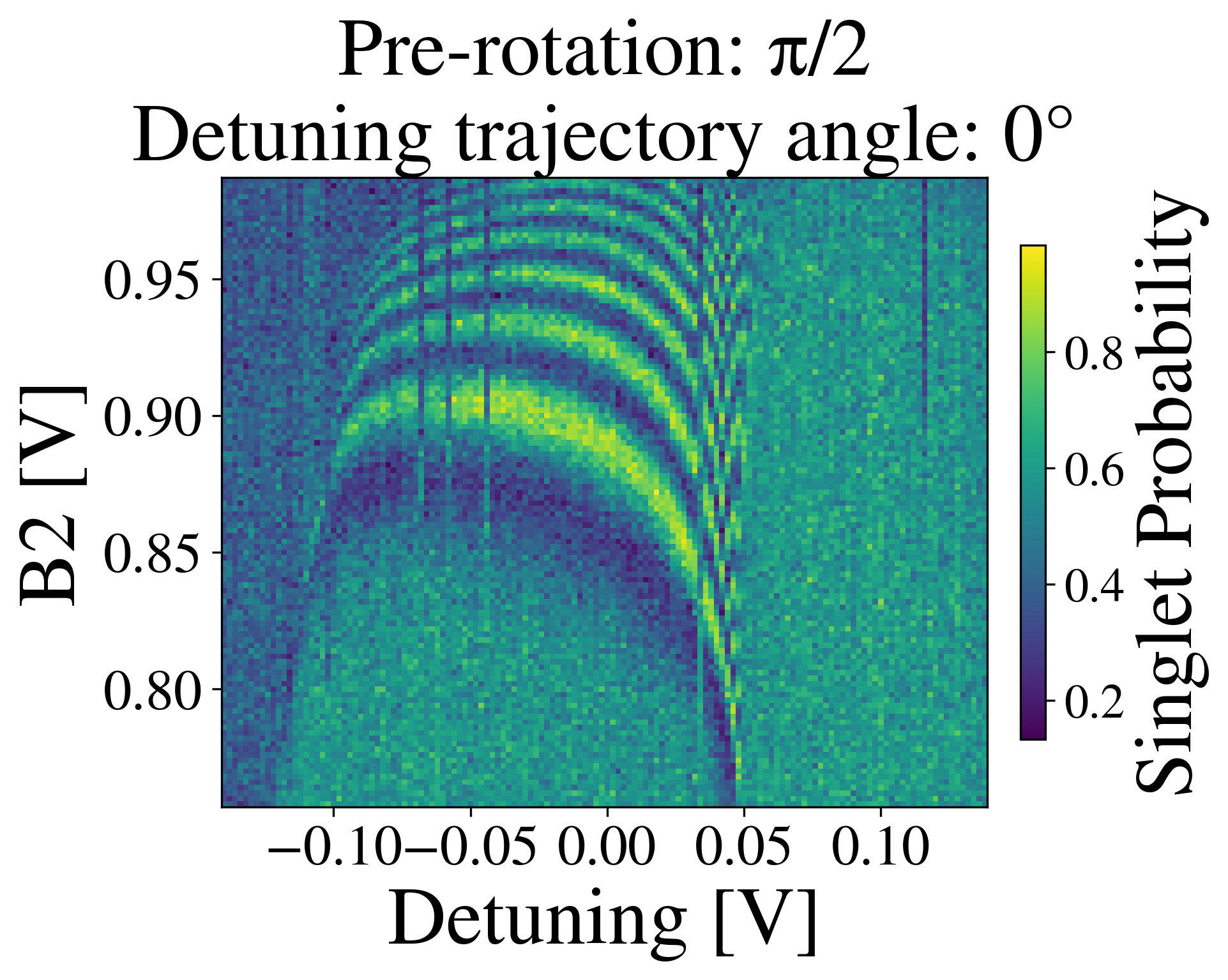}};
\node[right=of img2] (img3) {\includegraphics[width=3.8cm]{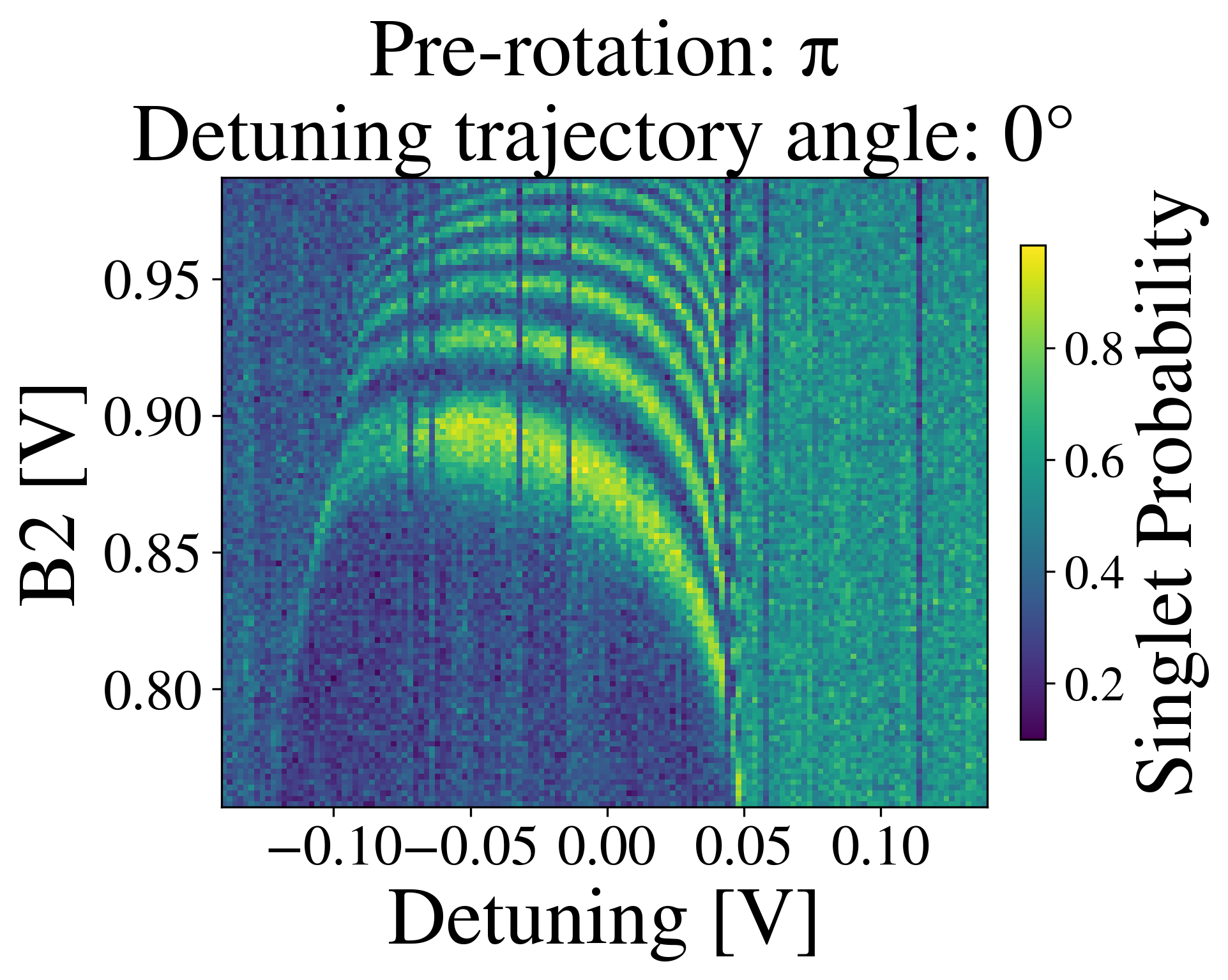}};
\node[right=of img3] (img4) {\includegraphics[width=3.8cm]{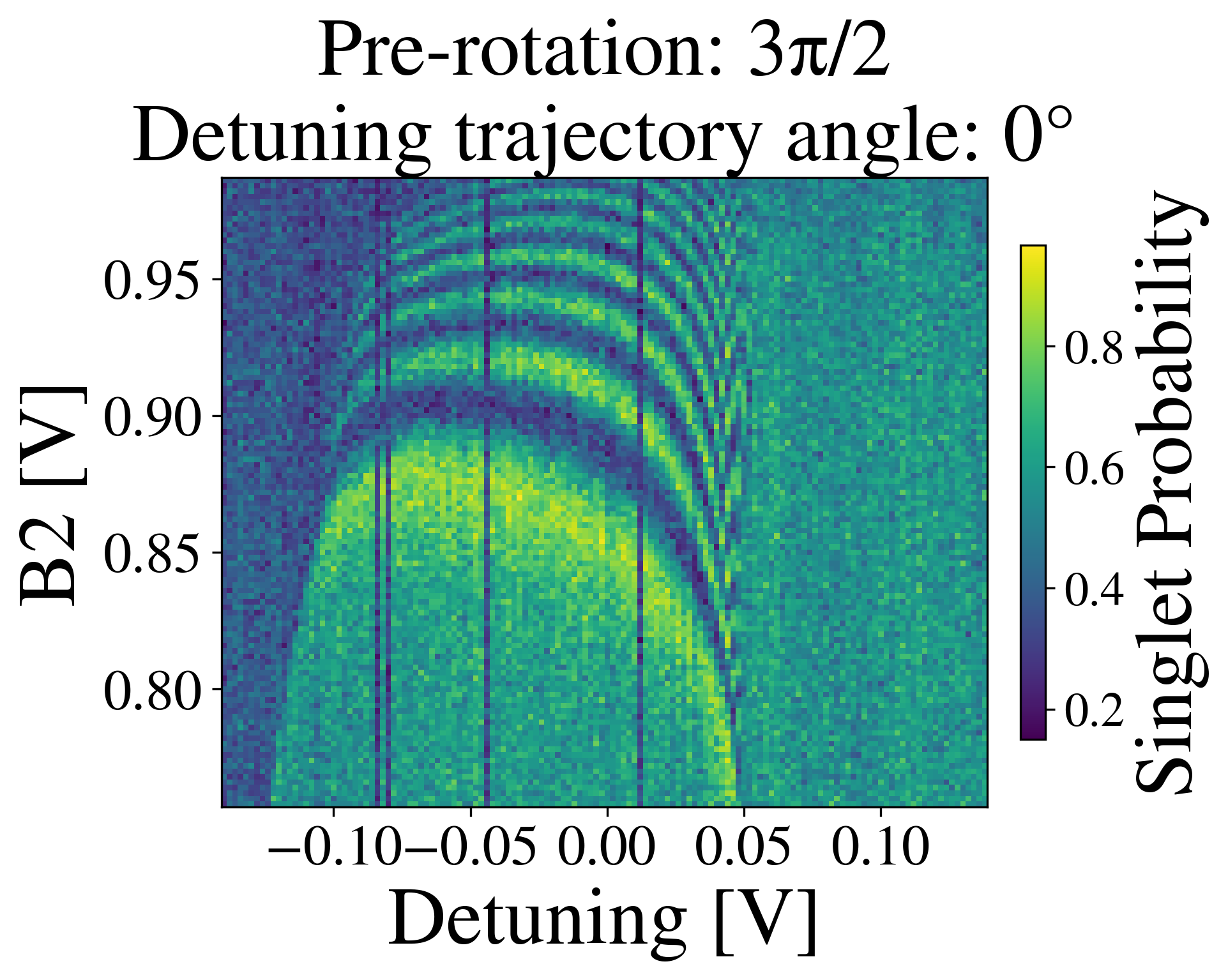}};

\node[below=of img1, xshift=1cm] (img5) {\includegraphics[height=5cm]{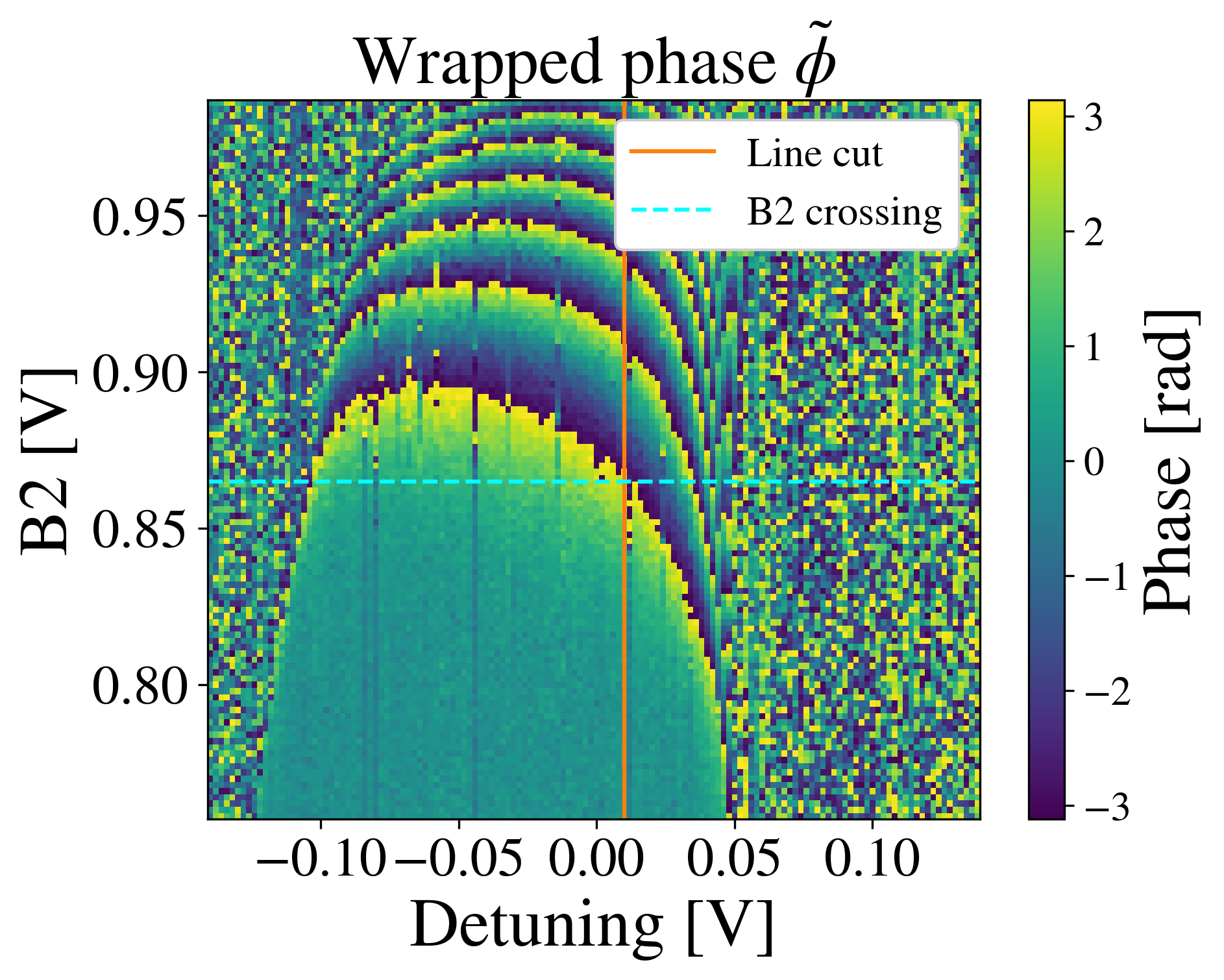}};
\node[below=of img3, xshift=2cm] (img6) {\includegraphics[height=5cm]{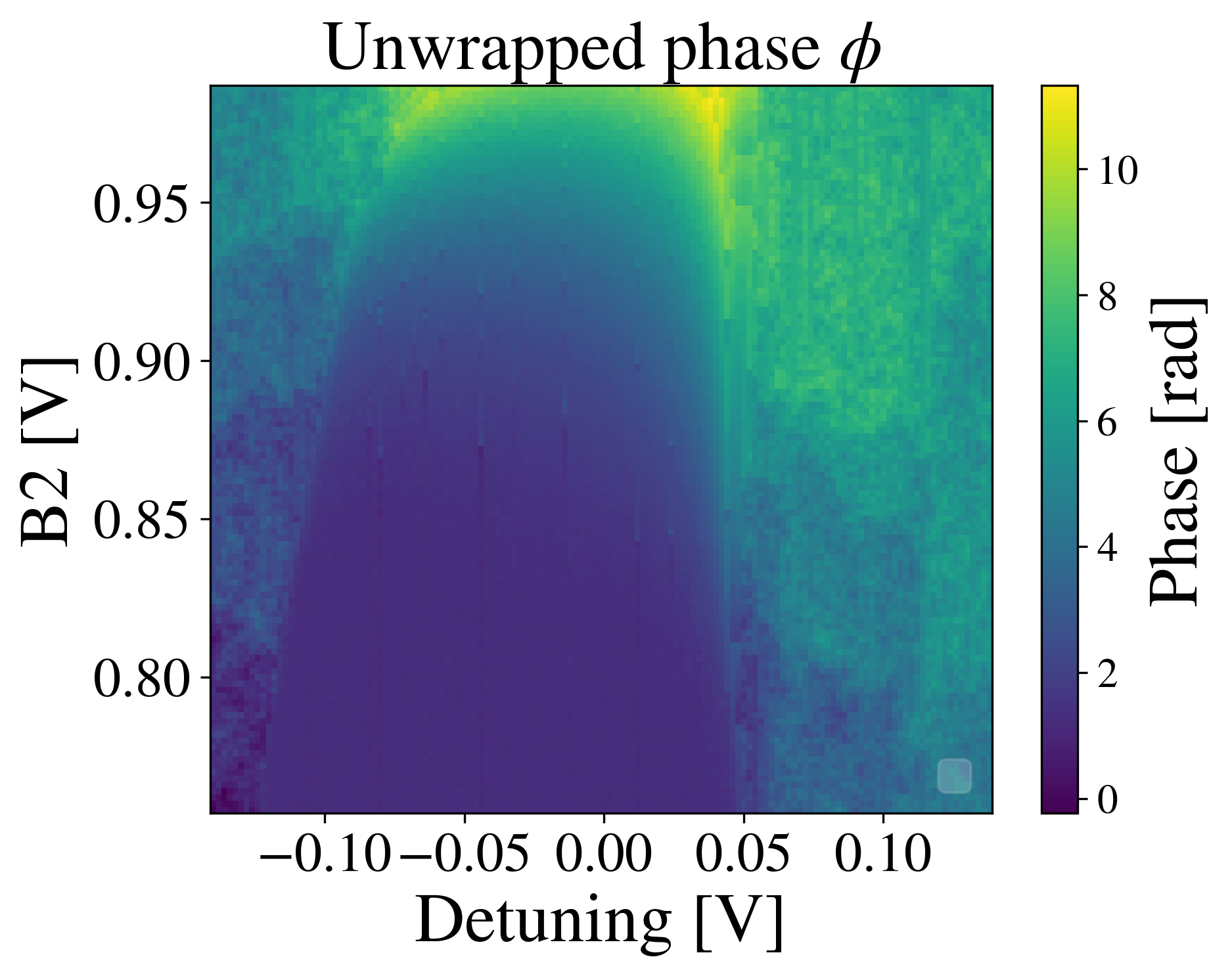}};
\node[below=of img5] (img7) {\includegraphics[height=5cm]{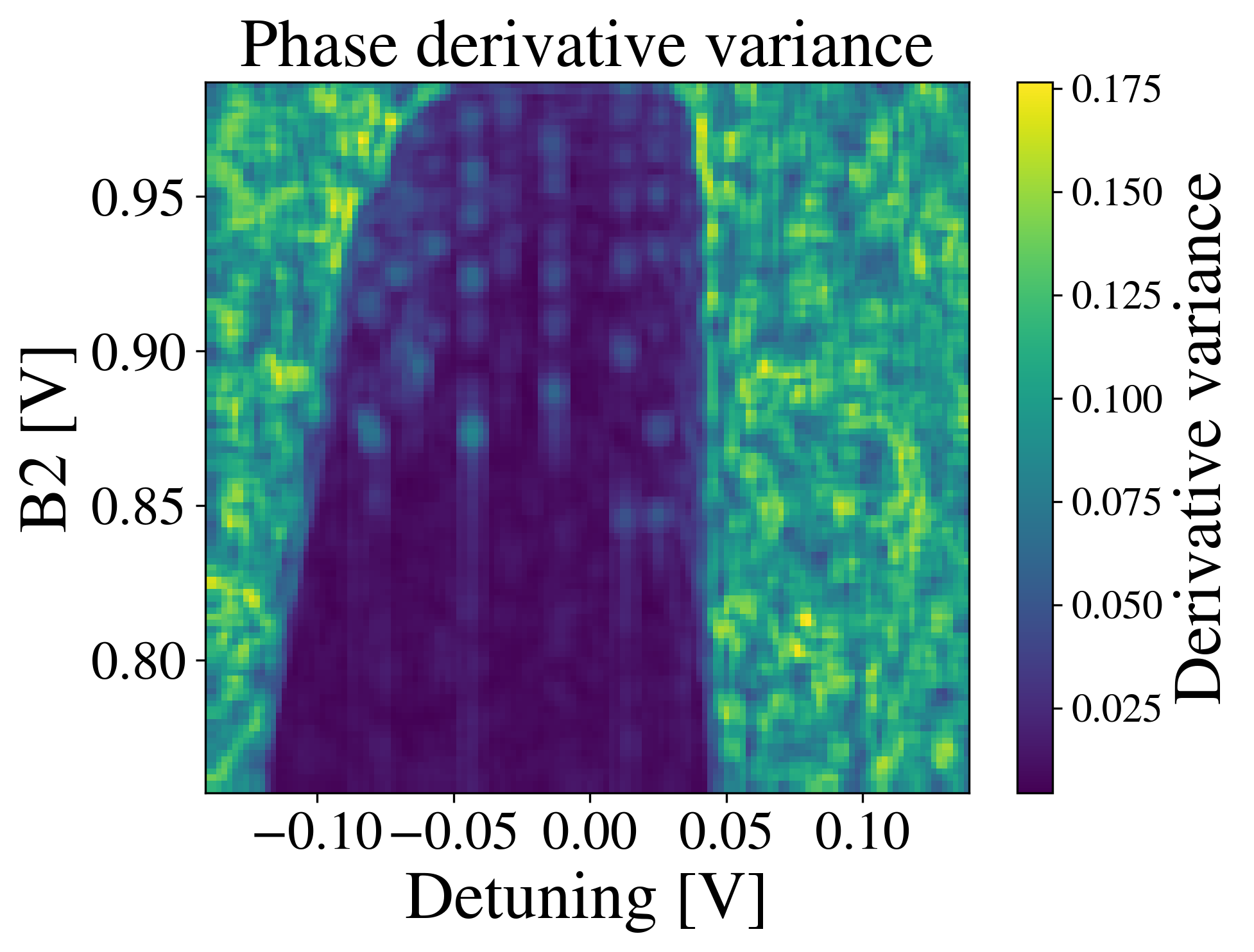}};
\node[below=of img6] (img8) {\includegraphics[height=5cm]{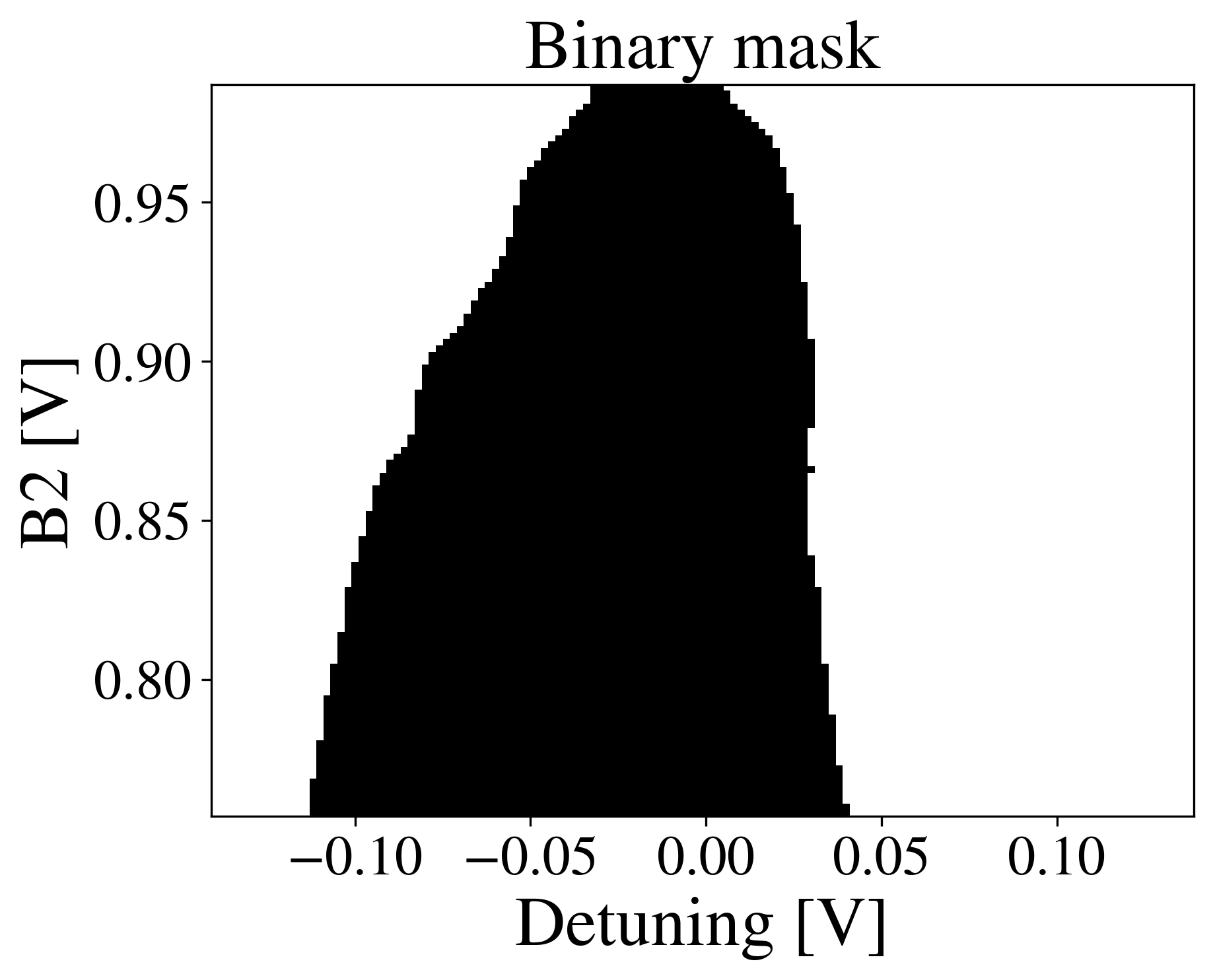}};
\node[below=of img7] (img9) {\includegraphics[height=5cm]{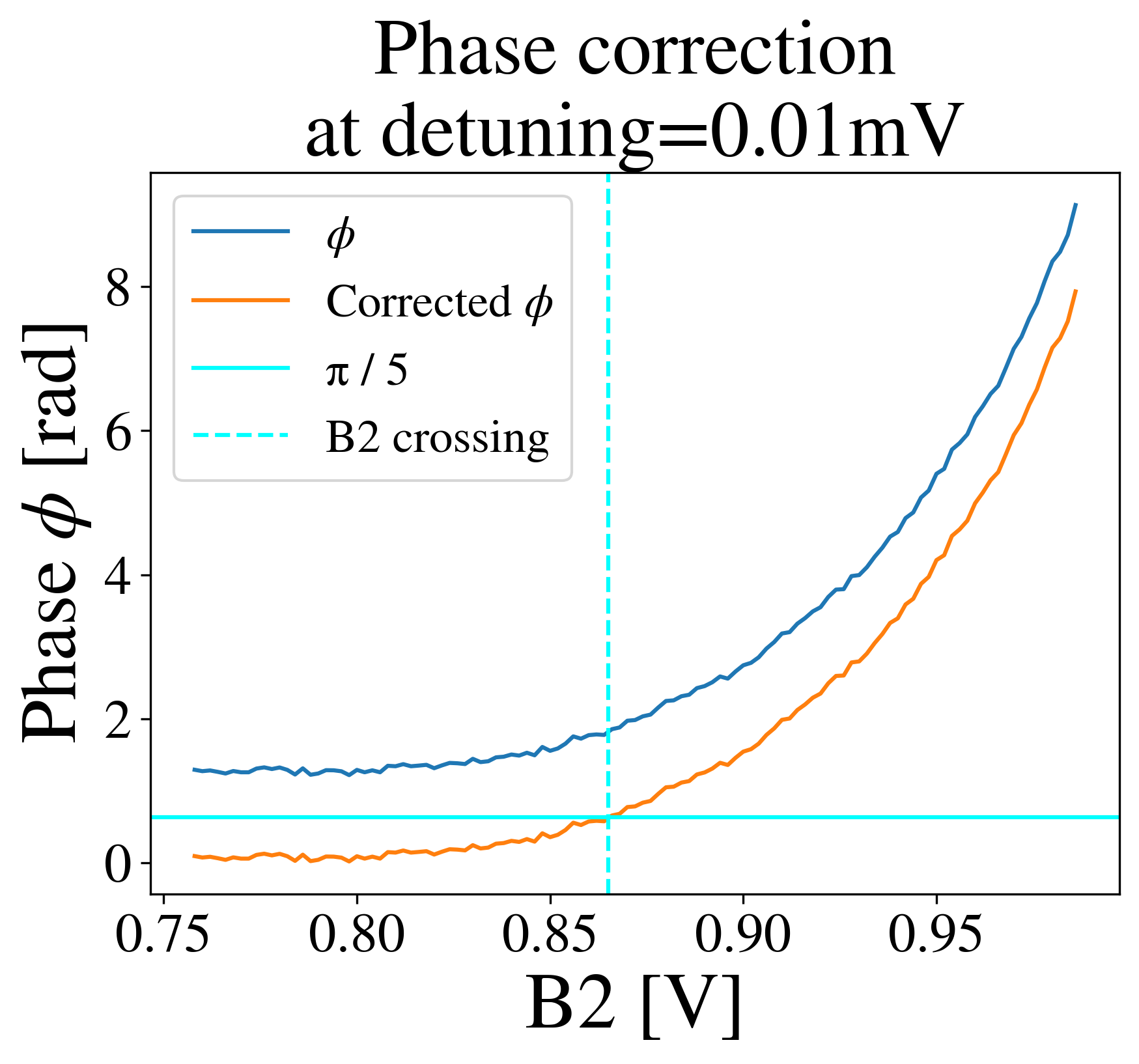}};
\node[below=of img8] (img10) {\includegraphics[height=5cm]{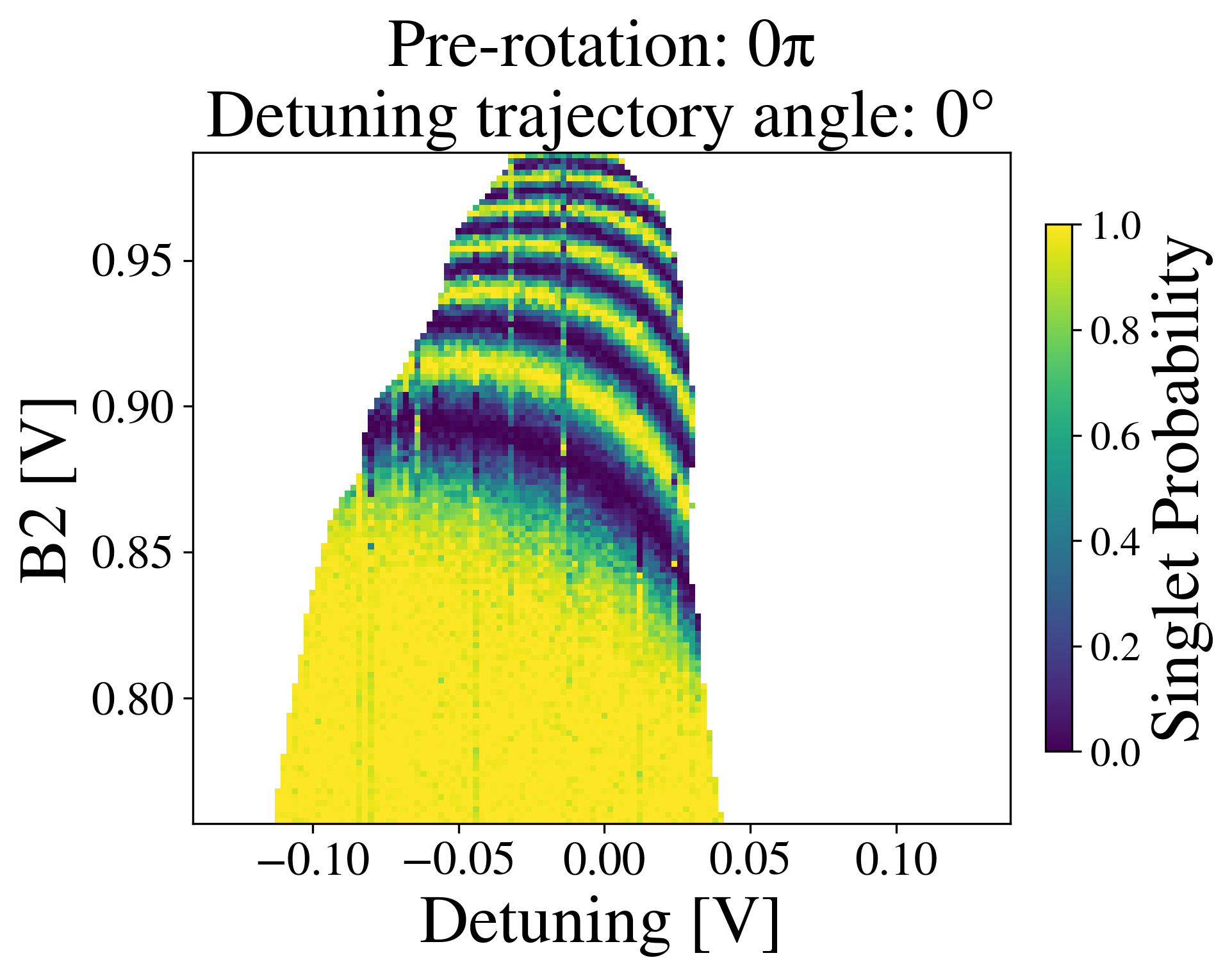}};

\draw[ultra thick] ($(img4.south)$) -- ($(img4.south)-(0,0.1)$);
\draw[ultra thick] ($(img4.south)-(0,0.1)$) -- ($(img5.north)-(0,0.1)$) node[pos=0.6,below] {Eq.~\ref{eq:wrapped_phase}};
\draw[->, ultra thick] ($(img5.north)-(0,0.1)$) -- ($(img5.north)-(0,0.3)$);

\draw[->, ultra thick] ($(img5.east)$) -- ($(img6.west)$) node[pos=0.5,below] {PUMA};

\draw[ultra thick] ($(img6.south)+(0,0.05)$) -- ($(img6.south)-(0,0.05)$);
\draw[ultra thick] ($(img6.south)-(0,0.05)$) -- ($(img7.north)-(0,0.05)$) node[pos=0.5,below] {PDV; Eq.~\ref{eq:pdv}};
\draw[->, ultra thick] ($(img7.north)-(0,0.05)$) -- ($(img7.north)-(0,0.25)$);

\draw[->, ultra thick] ($(img7.east)$) -- ($(img8.west)$) node[pos=0.5,below] {Threshold};

\draw[ultra thick] ($(img8.south)+(0,0.05)$) -- ($(img8.south)-(0,0.05)$);
\draw[ultra thick] ($(img8.south)-(0,0.05)$) -- ($(img9.north)-(0,0.05)$) node[pos=0.5,below] {Zero phase};
\draw[->, ultra thick] ($(img9.north)-(0,0.05)$) -- ($(img9.north)-(0,0.25)$);

\draw[->, ultra thick] ($(img9.east)$) -- ($(img10.west)$) node[pos=0.5,below] {Mask};

\draw[thick] (img1.north west) rectangle (img4.south east);

\end{tikzpicture}
\caption{\textbf{Data processing pipeline for $0\degree$ detuning.}
The pipeline steps leading up to a mask applied to the wrapped phase. We use the masks on the unwrapped phases to fit our model to the full 3D data.
}
\label{fig:data_processing_pipeline}
\end{figure*}

\noindent
Our approach to analyzing the collection of fingerprints is to extract the phase for each slice separately and then combine the resulting high quality phase information into a three-dimensional dataset to obtain a fitted three-dimensional model. The full data processing pipeline is presented in Figure~\ref{fig:data_processing_pipeline}. 
For these singlet-prep, $J_n$-rotation fingerprints, the ideal measurement signal for the singlet return probability is ${\rm PS}(x, y) = 5/8 + 3/8 \cos \left[\phi(x, y)\right]$. More realistically, this expression is written in the general form
\begin{equation}
d_i(x, y) = A(x, y) + B(x, y) \cos\left[\phi(x, y) + i\right] \quad ,
\end{equation}
where $i$ indicates a possible phase-shifted version of the experimental scan. As discussed, for each fingerprint we take the four phase-shifted scans, shifted by $0\pi$, $\pi / 2$, $\pi$, and $3\pi/2$, and this allows us to extract the 2D plane of wrapped phase via
\begin{equation}
\widetilde{\phi} = \tan^{-1}\left(\frac{d_{3\pi/2} - d_{\pi/2}}{d_{0} - d_{\pi}}\right) \quad ,
\label{eq:wrapped_phase}
\end{equation}
where $d_i$ corresponds to the $i$-phase-shifted data. We can then proceed to applying a phase unwrapping algorithm.

\subsection{Wrapped phase extraction, unwrapping, and high-confidence region segmentation}
\noindent
After applying Eq.~\ref{eq:wrapped_phase} to combine the scans into the 2D wrapped phase $\widetilde{\phi}$, we use the phase-unwrapping-max-flow (PUMA)~\cite{Bioucas-Dias2007} algorithm to carry out the phase unwrapping to obtain $\phi$ in that 2D slice.  PUMA converts $\widetilde{\phi}$ into a graph and estimates $\phi$ by solving an energy minimization problem on the graph. Due to the integral nature of the wrapping of the phase, the minimization can be performed through a sequence of binary optimizations -- that is, by making binary adjustments to get closer to the integer increments that account for the phase wrapping. Each binary optimization can be efficiently computed by a max-flow/min-cut calculation, as described in~\cite{Bioucas-Dias2007}. We manually spot-check the algorithm along line cuts of the wrapped phase data and determine that the algorithm unwraps the phase correctly, matching the expected increments of $2 \pi$. We do need to shift the extracted phase to correct for zero phase, but first we find the region of high-confidence phase values.

We estimate the region of ``high-confidence'' phase by using the phase derivative variance (PDV) (see~\cite{Yu2019} for example); however, we apply the PDV to the unwrapped phase, as opposed to the typical use case of applying the PDV to the wrapped phase:

{\fontsize{8pt}{9pt}\selectfont
\begin{equation}
{\rm pdv}_{p,q} = \frac{\sqrt{\sum_{i,j} \left(\Delta_{i,j}^{x} - \overbar{\Delta_{p,q}^{x}}\right)^2} + \sqrt{\sum_{i,j} \left(\Delta_{i,j}^{y} - \overbar{\Delta_{p,q}^{y}}\right)^2}}{N^2}
\label{eq:pdv}
\end{equation}
}
where $i,j$ are summed over a $N\times N$ window centered at $p,q$, the $\Delta_{i,j}^{x,y}$ quantities are the partial $x,y$ derivatives of the unwrapped phase $\widetilde{\phi}$, and the $\overbar{\Delta_{p,q}^{x,y}}$ terms represent the average of these derivatives over the window.  The reason we apply the PDV to the unwrapped phase is that we are interested in extracting a region of high-confidence phase, as opposed to informing and guiding a phase unwrapping algorithm toward better results.

Our final step in the phase extraction is to segment the data based on the PDV. As we are not focused on optimizing image segmentation, we use a fairly heavy-handed approach: we threshold the PDV image at the 40th percentile to obtain a binary mask. Once we have this mask, we return to the question of phase offset. We shift the phase by the average value of the 5th percentile of the data to zero, understanding, from the nature of the scans, that much of the data is in the small phase region. The result can be seen in the final two figures of Figure~\ref{fig:data_processing_pipeline}.

\subsection{Phase model}
\begin{figure*}[ht!]
\centering
\begin{tabular}{c}
\subfloat[\label{fig:model-a}]{\includegraphics[width=140mm]{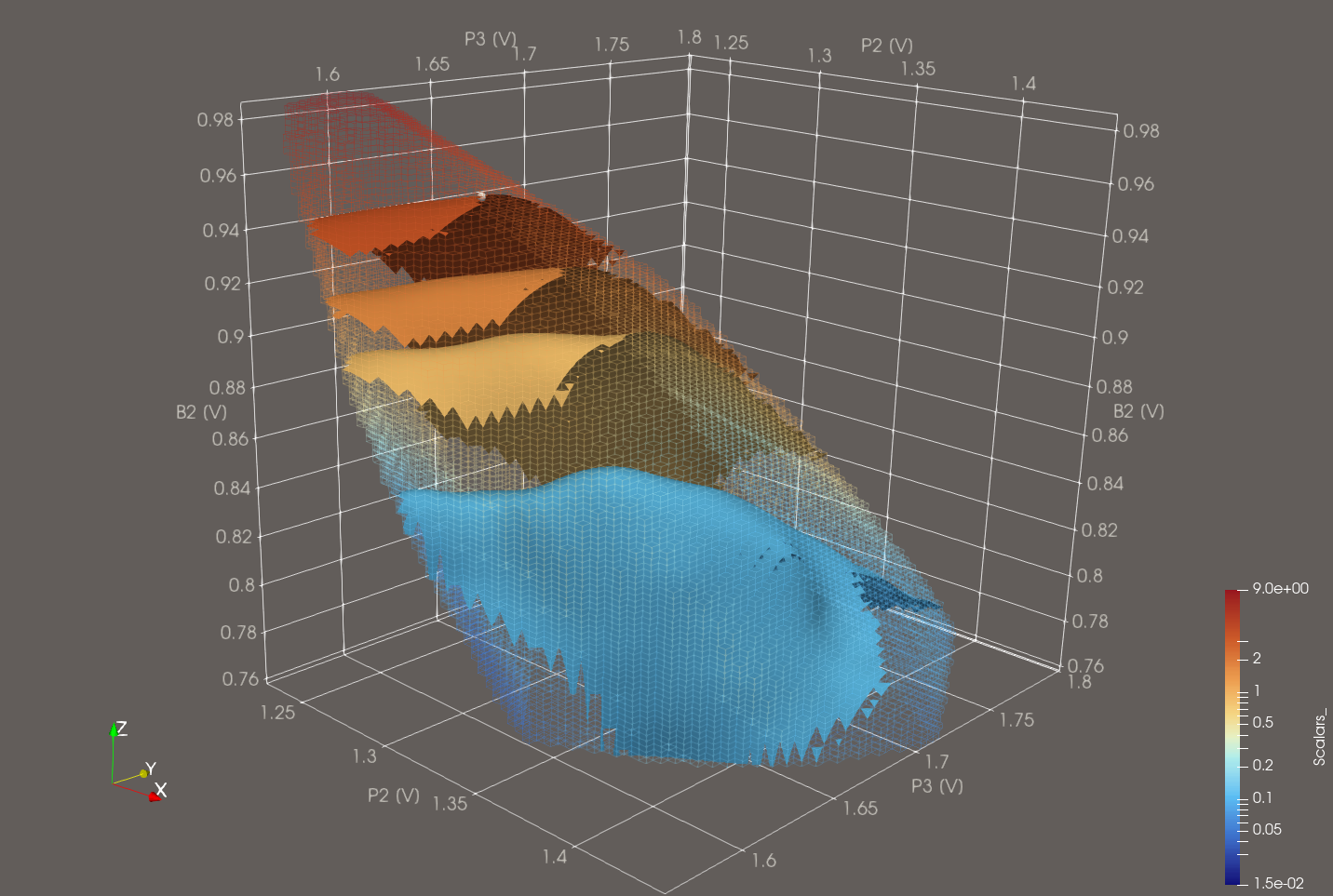}}
\\
\subfloat[
\label{fig:model-b}]{\includegraphics[width=160mm]{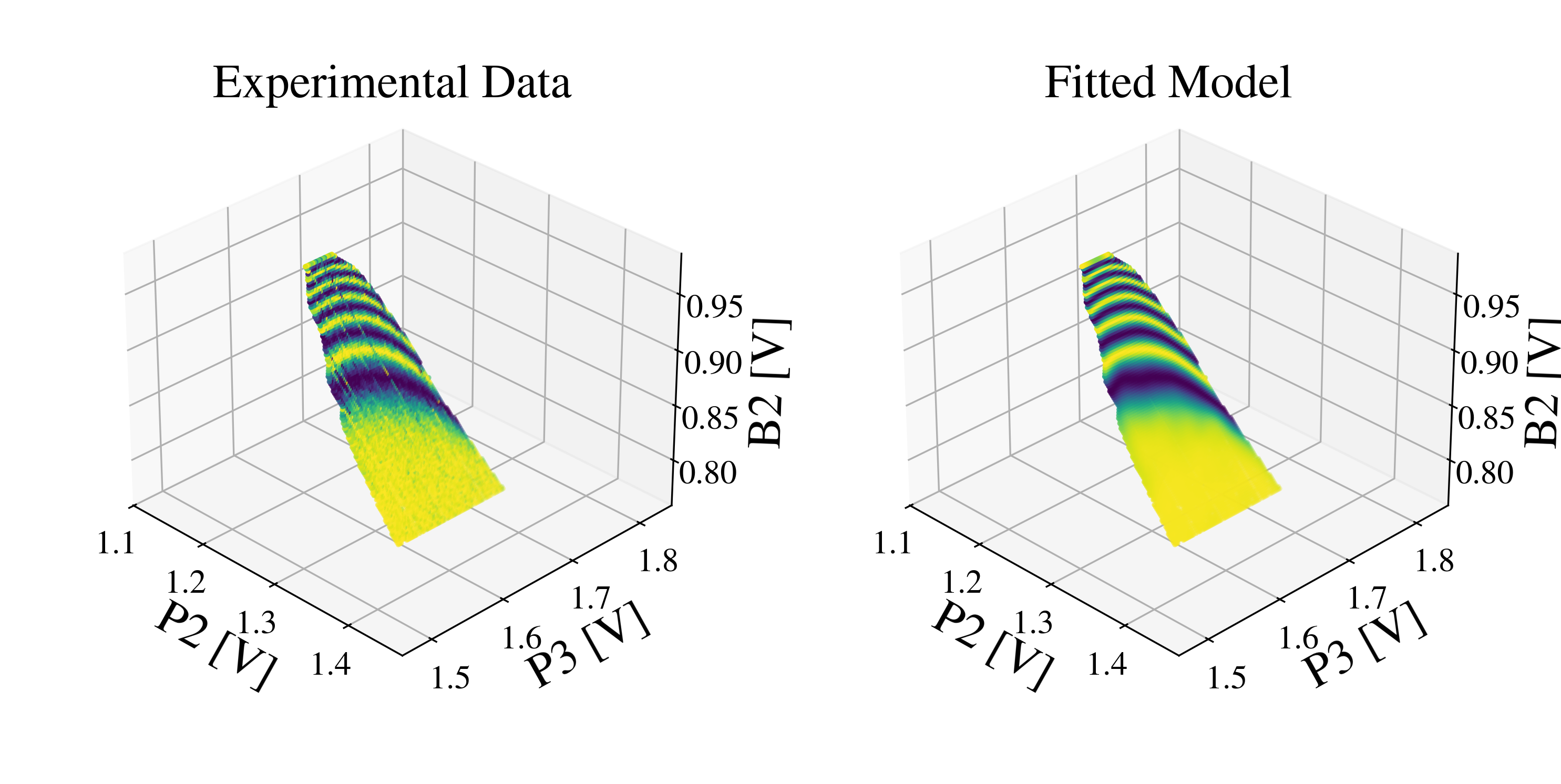}}
\end{tabular} 
\caption{
\textbf{Model.}
In (a), we show a rendering of the 3D phase spline model. The mesh outline indicates the convex hull of the confident phase region. From the bottom contour to top, the contours represent $0.1$, $\pi / 4$, $\pi / 2$, and $\pi$ radians. We have placed a small marble at the location of the symmetric operating point for a $\pi$ exchange pulse where phase gradients with respect to P2 and P3 vanish. In (b), we show a volumetric slice comparison of the experimental data and the spline model.
}
\label{fig:model}
\end{figure*}

\noindent
We build a three-dimensional model of the integrated exchange as a function of the three voltages $\phi(\mathbf{V}) = \phi(V_{\rm P2}, V_{\rm B2}, V_{\rm P3})$:
\begin{equation}
\phi(\mathbf{V}) = \frac{1}{\hbar N_{\rm pulses}} \int J(\mathbf{V}(t))\; {\rm d}t \quad ,
\label{eq:intj}
\end{equation}
where $N_{\rm pulses}$ is the number of pulses used to create the fingerprint.
It is useful intuitively to consider the case of ideal square wave voltage pulses occurring at the device, where we have nonzero exchange for time $T$ occurring for $N_{\rm pulses}$, and the phase $\phi$ is simply proportional to $J$ by a factor of $T/\hbar N_{\rm pulses}$. In this case we expect the phase $\phi$ to be exponential in $\mathbf{V}$~\cite{Burkard1999,Jirovec2025}. Though less direct of a relation in the non-ideal case, we treat the phase as having exponential character, and we fit a model to the logarithm of the phase, $\ln \phi$.

We attempt to fit the $\ln \phi$ data by first removing a 3mV radius cylindrical core enclosing data at the intersection of the planes, to remove any potential issues arising from that overlapping region of the data (this is indicated by a red circle in \Cref{fig:experimental_scans:c}). We then split off the top 10\% as a test set (largest 10\% of data B2 values). We perform a grid-search over polynomial models up to order 5 including varying levels of $L_1$ and $L_2$ regularizations, testing performance on the test set. We also manually search different spline models, testing performance against the test set. Using canned routines from \textsc{scipy}~\cite{2020SciPy-NMeth}, we find that a multiquadric radial basis function (RBF) interpolation (with degree 2 and smoothing 10) works quite well, reducing error on both training and test sets, outperforming polynomial fitting, evaluated based on root mean squared error (RMSE). Our best fitted model has a training set RMSE of $0.093$ radians and a test set RMSE of $0.164$ radians. Once we have optimized for the model and parameters using this train-test split approach, we fit the entire data (train plus test) to obtain the final spline model. In Figure~\ref{fig:model-a}, we show a \textsc{Paraview}~\cite{Ahrens2005,Ayachit2015} rendering of the fitted model's isophase contours enclosed in the convex hull of the confident phase region as well as the computed $\pi$ pulse point, optimizing for minimal $\partial \phi / \partial V_{P2,P3}$ gradients. (We optimize for the $\pi$-point by minimizing the following cost function $C = (\partial \phi / \partial V_{P2})^2 + (\partial \phi / \partial V_{P3})^2 + \lambda (\ln \phi - \ln \pi)^2$, where $\lambda$ is a penalty.) In Figure~\ref{fig:model-b}, we present a visual comparison of the experimental data and the fitted spline model on a single slice.

We note that more extensive model selection may be desirable, involving acquiring more data slices to be used as validation and test sets for cross-validation, model tuning, and model selection. We leave this for future work. 

\section{Conclusions}
\noindent
We have presented a measurement protocol and automatable data analysis pipeline to tomograph exchange phase in quantum dot devices. The protocol requires a series of phase-shifted, 2D experimental scans of barrier gate pulse amplitude ($V_{B2}$) against plunger gate detuning ($V_{P2}-V_{P3}$), rotating about a line of compensated pulsing in (P2, B2, P3) voltage space. In our demonstration, we acquired the 24 scans (4 phase shift conditions, 6 detuning angle conditions) of shape $115\times140$ pixels in about 21 hours. The analysis pipeline unwraps the phase and builds a model of the high-confidence phase volume in an automated fashion, running in about one minute. We applied these methods to characterize and model the phase in three-dimensional voltage space for a single axis (B2) in a 12-QD Intel Si/SiGe Tunnel Falls device.
The resulting model enables simulations of device operation to utilize realistic pulsing trajectories through phase space, incorporating experimental non-idealities such as skew; it enables intuitive experimental visualization of the exchange phase control space; it allows for pulse trajectory exploration or optimization, such as finding a pulse trajectory that minimizes $\partial \phi / \partial V_{P2,P3}$ gradients (in this work we optimized over the model to find an optimal $\pi$ pulse point) to reduce sensitivity to charge noise~\cite{Reed2016}; and it offers a point of connection to microscopic modeling, in that we may compare simulated phase from atomistic models incorporating disorder with the phase voltage dependence of measured devices. We anticipate using these techniques to rapidly model many exchange axes in other devices and tunings, and to investigate the physical origins of unusually-shaped exchange fingerprints.

We see the possibility for exploration and improvement in all aspects of the presented protocol. 
In particular, we believe improvements can be made to the phase segmentation/thresholding procedures, as well as those for model selection and fitting -- for example, collecting slices specifically intended for validation and testing.
Other future work could involve development of less intuitive voltage space slicing for more sparse and time-efficient data collection. 
Additionally, we are interested in how this method can help determine the signal transfer function unique to each voltage pulsing line in an experimental setup, which dictates the form of applied voltage pulses at the device.

The developments presented in this work provide an avenue for understanding the origins of device variability affecting device yield, and may inform the device fabrication process. Insights from this measurement protocol enable more accurate modeling of specific devices in their operating environment, which may facilitate improved error attribution through improved representation of underlying error sources. These developments also enable a systematic optimization of device control, which would be beneficial for maximizing device performance. We are keen to observe the potential impacts of the methods presented here on other qubit platforms.

\section{Acknowledgments}
\noindent
We acknowledge support from Intel Corporation for providing the device and Mayer Feldman, Matthew Curry, and Nathan Bishop for helpful discussions on device operation and performance. 
We thank Natalie D. Foster, Jacob D. Henshaw, Charlotte I. Evans, and Rohith Vudatha for offering advice on measurement and acquisition software usage. 

Research was sponsored by the Army Research Office and was accomplished under
Cooperative Agreement Number W911NF-22-2-0037. The views and conclusions contained in this document are those of the authors and should not be interpreted as representing the official policies, either expressed or implied, of the Army Research Office or the U.S. Government. The U.S. Government is authorized to reproduce and distribute reprints for Government purposes notwithstanding any copyright notation herein.

This article has been authored by an employee of National Technology \& Engineering Solutions of Sandia, LLC under Contract No. DE-NA0003525 with the U.S. Department of Energy (DOE). The employee owns all right, title and interest in and to the article and is solely responsible for its contents. The United States Government retains and the publisher, by accepting the article for publication, acknowledges that the United States Government retains a non-exclusive, paid-up, irrevocable, world-wide license to publish or reproduce the published form of this article or allow others to do so, for United States Government purposes. The DOE will provide public access to these results of federally sponsored research in accordance with the DOE Public Access Plan \url{https://www.energy.gov/downloads/doe-public-access-plan}.

This paper describes objective technical results and analysis. Any subjective views or opinions that might be expressed in the paper do not necessarily represent the views of the U.S. Department of Energy or the United States Government.

Sandia National Laboratories is a multimission laboratory managed and operated by National Technology \& Engineering Solutions of Sandia, LLC, a wholly owned subsidiary of Honeywell International Inc., for the U.S. Department of Energy’s National Nuclear Security Administration under contract DE-NA0003525. SAND\#0000-XXXXX

\appendix

\section{Device drift}
\label{sec:supplement}
\begin{figure*}[!h]
	\centering
	\includegraphics[width=0.7\textwidth]
    {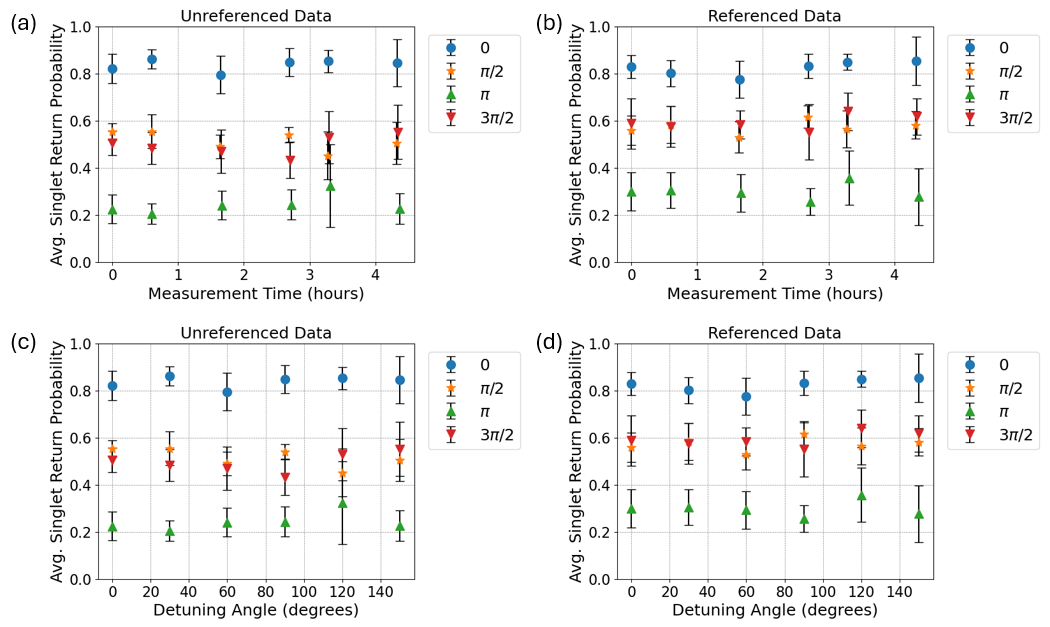}
	\caption{\textbf{Time-dependence of phase shift operations on the prepared singlet state.} Phase shift outcomes are calculated based on the singlet return probability for 11 random idle coordinates, defined by $V_{B2}=$0 V and $\lvert V_{P2} - V_{P3}\rvert < $ 14 mV, in each fingerprint measurement. Error bars represent standard deviation. To partially correct for drift in the SET output signal, SET reference measurements are recorded during qubit reset, and the difference in SET output during reset and measurement is pre-calibrated and assumed to be constant. (a) Unreferenced and (b) Referenced measurements of each phase shift outcome, as a function of time since the first measurement in a series for a given phase shift condition. (c) Unreferenced and (d) Referenced measurements of each phase shift outcome as a function of detuning angle condition.   }
    \label{fig:Drift}
\end{figure*}

\begin{figure*}[ht!]
\centering
\begin{tabular}{cc}
\subfloat[\label{fig:hires_experimental_scans:a}]{\includegraphics[width=70mm]{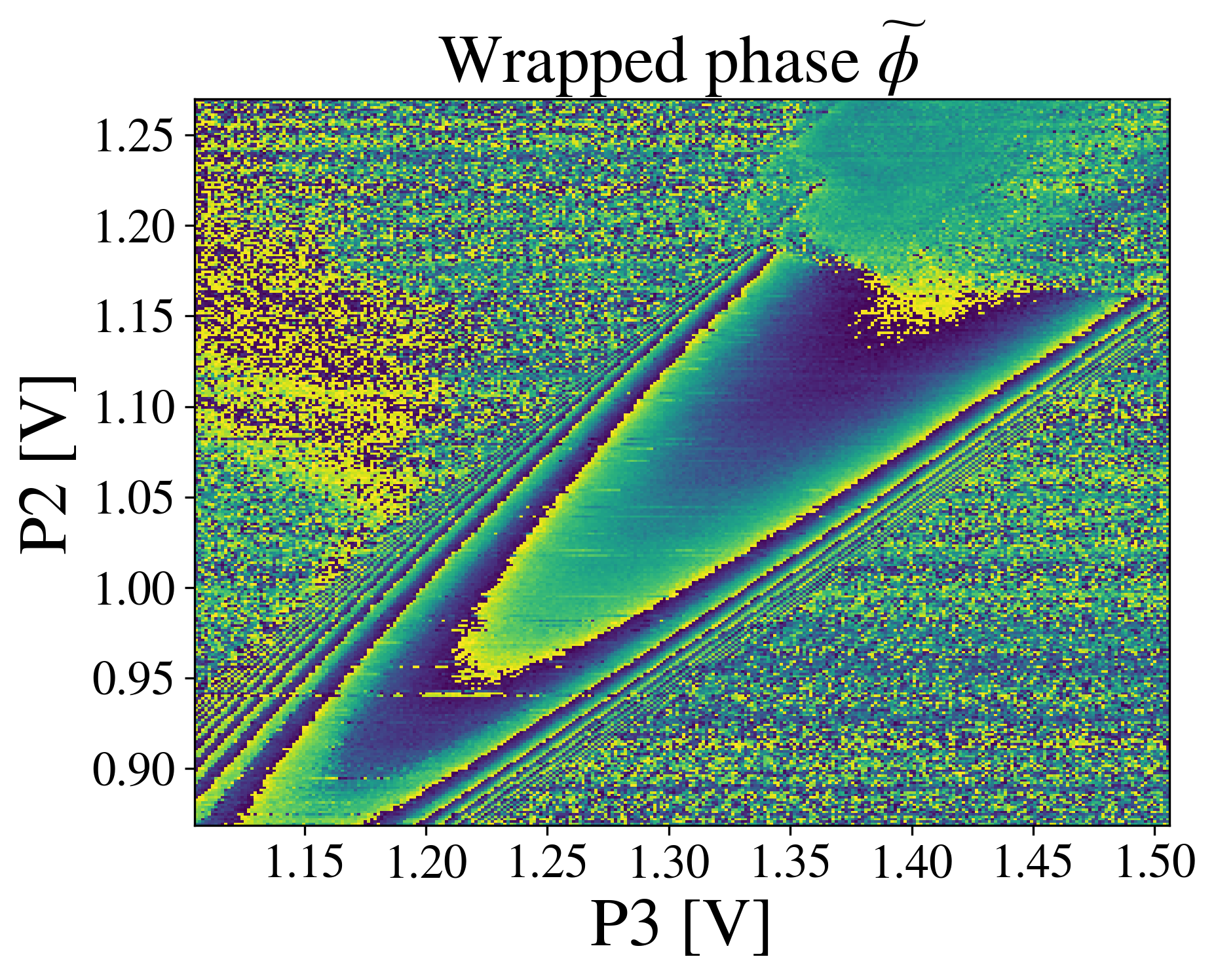}} &
\subfloat[\label{fig:hires_experimental_scans:b}]{
\includegraphics[width=70mm]{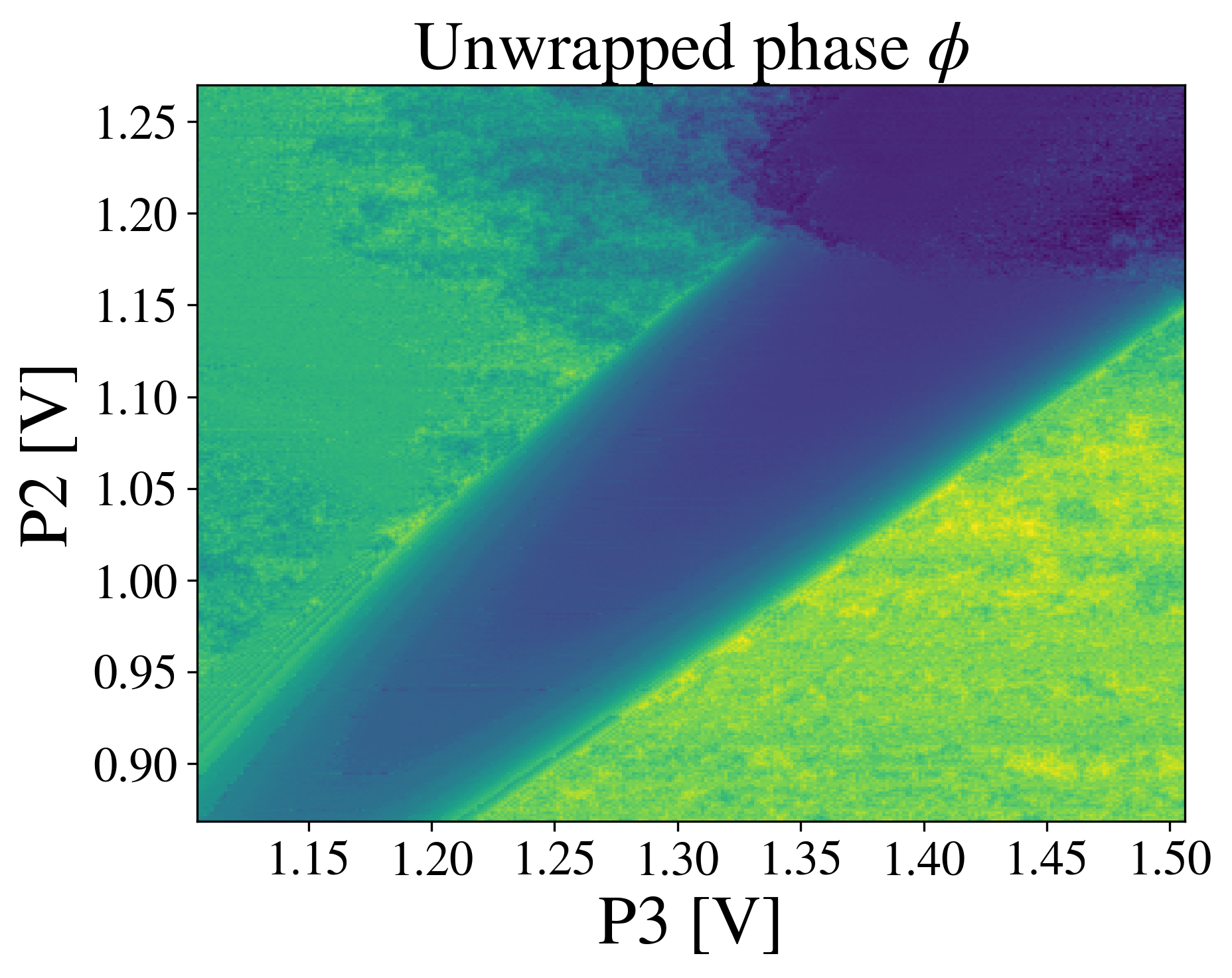}}
\end{tabular} 
\caption{
\textbf{High-resolution phase rasters.}
We show the successful phase unwrapping of high resolution scans, $300\times300$ pixels. This demonstrates that drift has not corrupted the phase extraction due to drift over $\sim 12$ hrs.
}
\label{fig:hires_experimental_scans}
\end{figure*}

\noindent
In our fingerprint scans, the voltage pulses applied during measurements have durations of $t_{\textrm{evol}}$ = 10 ns followed by buffer periods of $t_{\textrm{idle}}$ = 30 ns, making the total experiment duration (270 ns) between state preparation and measurement well within the $T_2^*$ dephasing time in our device, which is in excess of 1 $\mu$s~\cite{Madzik202512-spin-qubit, Neyens2024}. 
For scans of $115\times140$ pixels considered in this work, we find that our total data-collection time is approximately 21 hours.
To ensure that significant voltage drift did not occur in the device calibration during the time between phase-shifted scan variants, we evaluate the outcomes of each phase shift operation over time and find that they are consistent within error as shown in Fig. \ref{fig:Drift}. 
Drift in the SET output signal magnitude is compensated by a reference measurement during the qubit reset, and a pre-determined offset between SET outputs during reset and measurement which is assumed to be constant.
We also took higher-resolution phase raster data, with $300\times 300$ pixels, resulting in much longer data acquisition time, and presumably more exposure to drift, and had no difficulty in phase extraction, as shown in Figure \ref{fig:hires_experimental_scans}.

\bibliographystyle{unsrt}
\bibliography{phase_unwrapping}

@article{Reed2016,
  title={Reduced sensitivity to charge noise in semiconductor spin qubits via symmetric operation},
  author={Reed, MD and Maune, BM and Andrews, RW and Borselli, MG and Eng, K and Jura, MP and Kiselev, AA and Ladd, TD and Merkel, ST and Milosavljevic, I and others},
  journal={Physical review letters},
  volume={116},
  number={11},
  pages={110402},
  year={2016},
  publisher={APS}
}

@article{Ha2022,
author={Ha, Wonill
and Ha, Sieu D.
and Choi, Maxwell D.
and Tang, Yan
and Schmitz, Adele E.
and Levendorf, Mark P.
and Lee, Kangmu
and Chappell, James M.
and Adams, Tower S.
and Hulbert, Daniel R.
and Acuna, Edwin
and Noah, Ramsey S.
and Matten, Justine W.
and Jura, Michael P.
and Wright, Jeffrey A.
and Rakher, Matthew T.
and Borselli, Matthew G.},
title={A Flexible Design Platform for Si/SiGe Exchange-Only Qubits with Low Disorder},
journal={Nano Letters},
year={2022},
month={Feb},
day={09},
publisher={American Chemical Society},
volume={22},
number={3},
pages={1443-1448},
issn={1530-6984},
doi={10.1021/acs.nanolett.1c03026},
url={https://doi.org/10.1021/acs.nanolett.1c03026}
}

@article{deFuentes2026,
  title = {Running a Six-Qubit Quantum Circuit on a Silicon Spin-Qubit Array},
  author = {Fern\'andez de Fuentes, I. and Raymenants, E. and Undseth, B. and Pietx-Casas, O. and Philips, S. and M\k{a}dzik, M. and de Snoo, S.L. and Amitonov, S.V. and Tryputen, L. and Schmitz, A.T. and Matsuura, A.Y. and Scappucci, G. and Vandersypen, L.M.K.},
  journal = {PRX Quantum},
  volume = {7},
  issue = {1},
  pages = {010308},
  numpages = {13},
  year = {2026},
  month = {Jan},
  publisher = {American Physical Society},
  doi = {10.1103/f285-l2v5},
  url = {https://link.aps.org/doi/10.1103/f285-l2v5}
}

@article{DiVincenzo2000,
author={DiVincenzo, D. P.
and Bacon, D.
and Kempe, J.
and Burkard, G.
and Whaley, K. B.},
title={Universal quantum computation with the exchange interaction},
journal={Nature},
year={2000},
month={Nov},
day={01},
volume={408},
number={6810},
pages={339-342},
issn={1476-4687},
doi={10.1038/35042541},
url={https://doi.org/10.1038/35042541}
}

@misc{marcks2025valley,
      title={Valley Splitting Correlations Across a Silicon Quantum Well}, 
      author={Jonathan C. Marcks and Emily Eagen and Emma C. Brann and Merritt P. Losert and Tali Oh and John Reily and Christopher S. Wang and Daniel Keith and Fahd A. Mohiyaddin and Florian Luthi and Matthew J. Curry and Jiefei Zhang and F. Joseph Heremans and Mark Friesen and Mark A. Eriksson},
      year={2025},
      eprint={2504.12455},
      archivePrefix={arXiv},
      primaryClass={cond-mat.mes-hall},
      url={https://arxiv.org/abs/2504.12455}, 
}

@article{Mądzik2025,
author={M{\k{a}}dzik, Mateusz T.
and Luthi, Florian
and Guerreschi, Gian Giacomo
and Mohiyaddin, Fahd A.
and Borjans, Felix
and Chadwick, Jason D.
and Curry, Matthew J.
and Ziegler, Joshua
and Atanasov, Sarah
and Bavdaz, Peter L.
and Connors, Elliot J.
and Corrigan, J.
and Ercan, H. Ekmel
and Flory, Robert
and George, Hubert C.
and Harpt, Benjamin
and Henry, Eric
and Islam, Mohammad M.
and Khammassi, Nader
and Keith, Daniel
and Lampert, Lester F.
and Mladenov, Todor M.
and Morris, Randy W.
and Nethwewala, Aditi
and Neyens, Samuel
and Otten, Ren{\'e}
and Osuna Ibarra, Linda P.
and Patra, Bishnu
and Pillarisetty, Ravi
and Premaratne, Shavindra
and Ramsey, Mick
and Risinger, Andrew
and Rooney, John D.
and Savytskyy, Rostyslav
and Watson, Thomas F.
and Zietz, Otto K.
and Matsuura, Anne Y.
and Pellerano, Stefano
and Bishop, Nathaniel C.
and Roberts, Jeanette
and Clarke, James S.},
title={Operating two exchange-only qubits in parallel},
journal={Nature},
year={2025},
month={Nov},
day={01},
volume={647},
number={8091},
pages={870-875},
issn={1476-4687},
doi={10.1038/s41586-025-09767-5},
url={https://doi.org/10.1038/s41586-025-09767-5}
}

@article{Engel2001,
title = {Electron spins in quantum dots for spintronics and quantum computation},
journal = {Solid State Communications},
volume = {119},
number = {4},
pages = {229-236},
year = {2001},
issn = {0038-1098},
doi = {https://doi.org/10.1016/S0038-1098(01)00110-7},
url = {https://www.sciencedirect.com/science/article/pii/S0038109801001107},
author = {Hans-Andreas Engel and Patrik Recher and Daniel Loss}
}

@article{Koch2025,
author={Koch, Thomas
and Godfrin, Clement
and Adam, Viktor
and Ferrero, Julian
and Schroller, Daniel
and Glaeser, Noah
and Kubicek, Stefan
and Li, Ruoyu
and Loo, Roger
and Massar, Shana
and Simion, George
and Wan, Danny
and De Greve, Kristiaan
and Wernsdorfer, Wolfgang},
title={Industrial 300 mm wafer processed spin qubits in natural silicon/silicon-germanium},
journal={npj Quantum Information},
year={2025},
month={Apr},
day={05},
volume={11},
number={1},
pages={59},
issn={2056-6387},
doi={10.1038/s41534-025-01016-x},
url={https://doi.org/10.1038/s41534-025-01016-x}
}

@article{Steinacker2025,
author={Steinacker, Paul
and Dumoulin Stuyck, Nard
and Lim, Wee Han
and Tanttu, Tuomo
and Feng, MengKe
and Serrano, Santiago
and Nickl, Andreas
and Candido, Marco
and Cifuentes, Jesus D.
and Vahapoglu, Ensar
and Bartee, Samuel K.
and Hudson, Fay E.
and Chan, Kok Wai
and Kubicek, Stefan
and Jussot, Julien
and Canvel, Yann
and Beyne, Sofie
and Shimura, Yosuke
and Loo, Roger
and Godfrin, Clement
and Raes, Bart
and Baudot, Sylvain
and Wan, Danny
and Laucht, Arne
and Yang, Chih Hwan
and Saraiva, Andre
and Escott, Christopher C.
and De Greve, Kristiaan
and Dzurak, Andrew S.},
title={Industry-compatible silicon spin-qubit unit cells exceeding 99{\%} fidelity},
journal={Nature},
year={2025},
month={Oct},
day={01},
volume={646},
number={8083},
pages={81-87},
issn={1476-4687},
doi={10.1038/s41586-025-09531-9},
url={https://doi.org/10.1038/s41586-025-09531-9}
}

@article{Neyens2024,
author={Neyens, Samuel
and Zietz, Otto K.
and Watson, Thomas F.
and Luthi, Florian
and Nethwewala, Aditi
and George, Hubert C.
and Henry, Eric
and Islam, Mohammad
and Wagner, Andrew J.
and Borjans, Felix
and Connors, Elliot J.
and Corrigan, J.
and Curry, Matthew J.
and Keith, Daniel
and Kotlyar, Roza
and Lampert, Lester F.
and M{\k{a}}dzik, Mateusz T.
and Millard, Kent
and Mohiyaddin, Fahd A.
and Pellerano, Stefano
and Pillarisetty, Ravi
and Ramsey, Mick
and Savytskyy, Rostyslav
and Schaal, Simon
and Zheng, Guoji
and Ziegler, Joshua
and Bishop, Nathaniel C.
and Bojarski, Stephanie
and Roberts, Jeanette
and Clarke, James S.},
title={Probing single electrons across 300-mm spin qubit wafers},
journal={Nature},
year={2024},
month={May},
day={01},
volume={629},
number={8010},
pages={80-85},
issn={1476-4687},
doi={10.1038/s41586-024-07275-6},
url={https://doi.org/10.1038/s41586-024-07275-6}
}

@article{PRXQuantum.4.010329,
  title = {Complete Readout of Two-Electron Spin States in a Double Quantum Dot},
  author = {Nurizzo, Martin and Jadot, Baptiste and Mortemousque, Pierre-Andr\'e and Thiney, Vivien and Chanrion, Emmanuel and Niegemann, David and Dartiailh, Matthieu and Ludwig, Arne and Wieck, Andreas D. and B\"auerle, Christopher and Urdampilleta, Matias and Meunier, Tristan},
  journal = {PRX Quantum},
  volume = {4},
  issue = {1},
  pages = {010329},
  numpages = {11},
  year = {2023},
  month = {Mar},
  publisher = {American Physical Society},
  doi = {10.1103/PRXQuantum.4.010329},
  url = {https://link.aps.org/doi/10.1103/PRXQuantum.4.010329}
}

@article{PhysRevLett.116.110402,
  title = {Reduced Sensitivity to Charge Noise in Semiconductor Spin Qubits via Symmetric Operation},
  author = {Reed, M. D. and Maune, B. M. and Andrews, R. W. and Borselli, M. G. and Eng, K. and Jura, M. P. and Kiselev, A. A. and Ladd, T. D. and Merkel, S. T. and Milosavljevic, I. and Pritchett, E. J. and Rakher, M. T. and Ross, R. S. and Schmitz, A. E. and Smith, A. and Wright, J. A. and Gyure, M. F. and Hunter, A. T.},
  journal = {Phys. Rev. Lett.},
  volume = {116},
  issue = {11},
  pages = {110402},
  numpages = {6},
  year = {2016},
  month = {Mar},
  publisher = {American Physical Society},
  doi = {10.1103/PhysRevLett.116.110402},
  url = {https://link.aps.org/doi/10.1103/PhysRevLett.116.110402}
}

@article{Andrews2019,
  title={Quantifying error and leakage in an encoded Si/SiGe triple-dot qubit},
  author={Andrews, R.W. and Jones, C. and Reed, M. D.},
  journal={Nat. Nanotechnol.},
  volume={14},
  year={2019}
}

@article{Barnes2020,
  title={Correcting Distortion of Base-band Exchange Pulses in Quantum Dot Qubits},
  author={Barnes, David},
  journal={Bulletin of the American Physical Society},
  volume={65},
  year={2020},
  publisher={APS}
}

@article{Fedele2021,
  title = {Simultaneous Operations in a Two-Dimensional Array of Singlet-Triplet Qubits},
  author = {Fedele, Federico and Chatterjee, Anasua and Fallahi, Saeed and Gardner, Geoffrey C. and Manfra, Michael J. and Kuemmeth, Ferdinand},
  journal = {PRX Quantum},
  volume = {2},
  issue = {4},
  pages = {040306},
  numpages = {12},
  year = {2021},
  month = {Oct},
  publisher = {American Physical Society},
  doi = {10.1103/PRXQuantum.2.040306},
  url = {https://link.aps.org/doi/10.1103/PRXQuantum.2.040306}
}

@article{Ladd2010,
author={Ladd, T. D.
and Jelezko, F.
and Laflamme, R.
and Nakamura, Y.
and Monroe, C.
and O'Brien, J. L.},
title={Quantum computers},
journal={Nature},
year={2010},
month={Mar},
day={01},
volume={464},
number={7285},
pages={45-53},
issn={1476-4687},
doi={10.1038/nature08812},
url={https://doi.org/10.1038/nature08812}
}

@article{Ni2025,
doi = {10.1088/1674-1056/ad8db1},
url = {https://doi.org/10.1088/1674-1056/ad8db1},
year = {2025},
month = {jan},
publisher = {Chinese Physical Society and IOP Publishing Ltd},
volume = {34},
number = {1},
pages = {010308},
author = {Ni, Ming and Ma, Rong-Long and Kong, Zhen-Zhen and Chu, Ning and Liao, Wei-Zhu and Zhu, Sheng-Kai and Wang, Chu and Luo, Gang and Liu, Di and Cao, Gang and Wang, Gui-Lei and Li, Hai-Ou and Guo, Guo-Ping},
title = {Correcting on-chip distortion of control pulses with silicon spin qubits},
journal = {Chinese Physics B}
}

@article{Zhang2025,
author={Zhang, Xin
and Morozova, Elizaveta
and Rimbach-Russ, Maximilian
and Jirovec, Daniel
and Hsiao, Tzu-Kan
and Fari{\~{n}}a, Pablo Cova
and Wang, Chien-An
and Oosterhout, Stefan D.
and Sammak, Amir
and Scappucci, Giordano
and Veldhorst, Menno
and Vandersypen, Lieven M. K.},
title={Universal control of four singlet--triplet qubits},
journal={Nature Nanotechnology},
year={2025},
month={Feb},
day={01},
volume={20},
number={2},
pages={209-215},
issn={1748-3395},
doi={10.1038/s41565-024-01817-9},
url={https://doi.org/10.1038/s41565-024-01817-9}
}

@article{Madzik202512-spin-qubit,
author={George, Hubert C.
and M{\k{a}}dzik, Mateusz T.
and Henry, Eric M.
and Wagner, Andrew J.
and Islam, Mohammad M.
and Borjans, Felix
and Connors, Elliot J.
and Corrigan, J.
and Curry, Matthew
and Harper, Michael K.
and Keith, Daniel
and Lampert, Lester
and Luthi, Florian
and Mohiyaddin, Fahd A.
and Murcia, Sandra
and Nair, Rohit
and Nahm, Rambert
and Nethwewala, Aditi
and Neyens, Samuel
and Patra, Bishnu
and Raharjo, Roy D.
and Rogan, Carly
and Savytskyy, Rostyslav
and Watson, Thomas F.
and Ziegler, Josh
and Zietz, Otto K.
and Pellerano, Stefano
and Pillarisetty, Ravi
and Bishop, Nathaniel C.
and Bojarski, Stephanie A.
and Roberts, Jeanette
and Clarke, James S.},
title={12-Spin-Qubit Arrays Fabricated on a 300 mm Semiconductor Manufacturing Line},
journal={Nano Letters},
year={2025},
month={Jan},
day={15},
publisher={American Chemical Society},
volume={25},
number={2},
pages={793-799},
issn={1530-6984},
doi={10.1021/acs.nanolett.4c05205},
url={https://doi.org/10.1021/acs.nanolett.4c05205}
}

@article{Rol2020,
    author = {Rol, M. A. and Ciorciaro, L. and Malinowski, F. K. and Tarasinski, B. M. and Sagastizabal, R. E. and Bultink, C. C. and Salathe, Y. and Haandbaek, N. and Sedivy, J. and DiCarlo, L.},
    title = {Time-domain characterization and correction of on-chip distortion of control pulses in a quantum processor},
    journal = {Applied Physics Letters},
    volume = {116},
    number = {5},
    pages = {054001},
    year = {2020},
    month = {02},
    issn = {0003-6951},
    doi = {10.1063/1.5133894},
    url = {https://doi.org/10.1063/1.5133894},
    eprint = {https://pubs.aip.org/aip/apl/article-pdf/doi/10.1063/1.5133894/14531145/054001_1_online.pdf},
}

@article{Buterakos2021,
  title = {Spin-Valley Qubit Dynamics in Exchange-Coupled Silicon Quantum Dots},
  author = {Buterakos, Donovan and Das Sarma, Sankar},
  journal = {PRX Quantum},
  volume = {2},
  issue = {4},
  pages = {040358},
  numpages = {15},
  year = {2021},
  month = {Dec},
  publisher = {American Physical Society},
  doi = {10.1103/PRXQuantum.2.040358},
  url = {https://link.aps.org/doi/10.1103/PRXQuantum.2.040358}
}

@article{Tariq2022,
author={Tariq, Bilal
and Hu, Xuedong},
title={Impact of the valley orbit coupling on exchange gate for spin qubits in silicon},
journal={npj Quantum Information},
year={2022},
month={May},
day={10},
volume={8},
number={1},
pages={53},
issn={2056-6387},
doi={10.1038/s41534-022-00554-y},
url={https://doi.org/10.1038/s41534-022-00554-y}
}

@article{Madzik2025operating,
  title={Operating two exchange-only qubits in parallel},
  author={M{\k{a}}dzik, Mateusz T and Luthi, Florian and Guerreschi, Gian Giacomo and Mohiyaddin, Fahd A and Borjans, Felix and Chadwick, Jason D and Curry, Matthew J and Ziegler, Joshua and Atanasov, Sarah and Bavdaz, Peter L and others},
  journal={arXiv preprint arXiv:2504.01191},
  year={2025}
}

@inproceedings{Lanza2022,
  title={Quantum characterization of 6-dot exchange-only qubit arrays in the SLEDGE architecture},
  author={Lanza, Robert and Holman, Nathan},
  booktitle={APS March Meeting Abstracts},
  volume={2022},
  pages={Z39--012},
  year={2022}
}

@inproceedings{Acuna2022,
  title={Observation of anomalous T 1 relaxation during exchange in Si/SiGe spin qubits},
  author={Acuna, Edwin},
  booktitle={APS March Meeting Abstracts},
  volume={2022},
  pages={G39--007},
  year={2022}
}

@article{Burkard1999,
  title={Coupled quantum dots as quantum gates},
  author={Burkard, Guido and Loss, Daniel and DiVincenzo, David P},
  journal={Physical Review B},
  volume={59},
  number={3},
  pages={2070},
  year={1999},
  publisher={APS}
}

@article{Jirovec2025,
  title={Mitigation of exchange crosstalk in dense quantum dot arrays},
  author={Jirovec, Daniel and Fari{\~n}a, Pablo Cova and Reale, Stefano and Oosterhout, Stefan D and Zhang, Xin and de Snoo, Sander and Sammak, Amir and Scappucci, Giordano and Veldhorst, Menno and Vandersypen, Lieven MK},
  journal={Physical Review Applied},
  volume={24},
  number={3},
  pages={034051},
  year={2025},
  publisher={APS}
}

@article{Bioucas-Dias2007,
  author={Bioucas-Dias, José M. and Valadão, Gonçalo},
  journal={IEEE Transactions on Image Processing}, 
  title={Phase Unwrapping via Graph Cuts}, 
  year={2007},
  volume={16},
  number={3},
  pages={698-709},
  doi={10.1109/TIP.2006.888351},
}

@article{Judge1994,
title = {A review of phase unwrapping techniques in fringe analysis},
journal = {Optics and Lasers in Engineering},
volume = {21},
number = {4},
pages = {199-239},
year = {1994},
issn = {0143-8166},
doi = {https://doi.org/10.1016/0143-8166(94)90073-6},
url = {https://www.sciencedirect.com/science/article/pii/0143816694900736},
author = {T.R. Judge and P.J. Bryanston-Cross},
}

@article{Wang2022,
  title={Deep learning spatial phase unwrapping: a comparative review},
  author={Wang, Kaiqiang and Kemao, Qian and Di, Jianglei and Zhao, Jianlin},
  journal={Advanced Photonics Nexus},
  volume={1},
  number={1},
  pages={014001--014001},
  year={2022},
  publisher={Society of Photo-Optical Instrumentation Engineers}
}

@article{Yu2019,
  author={Yu, Hanwen and Lan, Yang and Yuan, Zhihui and Xu, Junyi and Lee, Hyongki},
  journal={IEEE Geoscience and Remote Sensing Magazine}, 
  title={Phase Unwrapping in InSAR : A Review}, 
  year={2019},
  volume={7},
  number={1},
  pages={40-58},
  doi={10.1109/MGRS.2018.2873644},
}

@article{Jenkinson2012,
  title={Fsl},
  author={Jenkinson, Mark and Beckmann, Christian F and Behrens, Timothy EJ and Woolrich, Mark W and Smith, Stephen M},
  journal={Neuroimage},
  volume={62},
  number={2},
  pages={782--790},
  year={2012},
  publisher={Elsevier}
}

@article{Jenkinson2003,
  title={Fast, automated, N-dimensional phase-unwrapping algorithm},
  author={Jenkinson, Mark},
  journal={Magnetic Resonance in Medicine: An Official Journal of the International Society for Magnetic Resonance in Medicine},
  volume={49},
  number={1},
  pages={193--197},
  year={2003},
  publisher={Wiley Online Library}
}

@article{Zuo2022,
  title={Deep learning in optical metrology: a review},
  author={Zuo, Chao and Qian, Jiaming and Feng, Shijie and Yin, Wei and Li, Yixuan and Fan, Pengfei and Han, Jing and Qian, Kemao and Chen, Qian},
  journal={Light: Science \& Applications},
  volume={11},
  number={1},
  pages={39},
  year={2022},
  publisher={Nature Publishing Group UK London}
}

@article{Aiello2007,
title = {Green's formulation for robust phase unwrapping in digital holography},
journal = {Optics and Lasers in Engineering},
volume = {45},
number = {6},
pages = {750-755},
year = {2007},
issn = {0143-8166},
doi = {https://doi.org/10.1016/j.optlaseng.2006.10.002},
url = {https://www.sciencedirect.com/science/article/pii/S014381660600193X},
author = {L. Aiello and D. Riccio and P. Ferraro and S. Grilli and L. Sansone and G. Coppola and S. {De Nicola} and A. Finizio},
}

@article{Feng2021,
title = {Calibration of fringe projection profilometry: A comparative review},
journal = {Optics and Lasers in Engineering},
volume = {143},
pages = {106622},
year = {2021},
issn = {0143-8166},
doi = {https://doi.org/10.1016/j.optlaseng.2021.106622},
url = {https://www.sciencedirect.com/science/article/pii/S0143816621000920},
author = {Shijie Feng and Chao Zuo and Liang Zhang and Tianyang Tao and Yan Hu and Wei Yin and Jiaming Qian and Qian Chen},
}

@article{Hu2020,
title = {Microscopic fringe projection profilometry: A review},
journal = {Optics and Lasers in Engineering},
volume = {135},
pages = {106192},
year = {2020},
issn = {0143-8166},
doi = {https://doi.org/10.1016/j.optlaseng.2020.106192},
url = {https://www.sciencedirect.com/science/article/pii/S0143816619319815},
author = {Yan Hu and Qian Chen and Shijie Feng and Chao Zuo},
}

@article{Su2001,
title = {Fourier transform profilometry:: a review},
journal = {Optics and Lasers in Engineering},
volume = {35},
number = {5},
pages = {263-284},
year = {2001},
issn = {0143-8166},
doi = {https://doi.org/10.1016/S0143-8166(01)00023-9},
url = {https://www.sciencedirect.com/science/article/pii/S0143816601000239},
author = {Xianyu Su and Wenjing Chen},
}

@article{Fong2022,
title = {Phase estimation via riesz transform in laser speckle interferometry for large-area damage imaging},
journal = {NDT \& E International},
volume = {132},
pages = {102711},
year = {2022},
issn = {0963-8695},
doi = {https://doi.org/10.1016/j.ndteint.2022.102711},
url = {https://www.sciencedirect.com/science/article/pii/S0963869522001104},
author = {Rey-Yie Fong and Fuh-Gwo Yuan},
}

@article{Tounsi2019,
author = {Yassine Tounsi and Manoj Kumar and Ahmed Siari and Fernando Mendoza-Santoyo and Abdelkrim Nassim and Osamu Matoba},
journal = {Opt. Lett.},
number = {14},
pages = {3434--3437},
publisher = {Optica Publishing Group},
title = {Digital four-step phase-shifting technique from a single fringe pattern using Riesz transform},
volume = {44},
month = {Jul},
year = {2019},
url = {https://opg.optica.org/ol/abstract.cfm?URI=ol-44-14-3434},
doi = {10.1364/OL.44.003434},
}

@article{Seelamantula2012,
author = {Chandra Sekhar Seelamantula and Nicolas Pavillon and Christian Depeursinge and Michael Unser},
journal = {J. Opt. Soc. Am. A},
number = {10},
pages = {2118--2129},
publisher = {Optica Publishing Group},
title = {Local demodulation of holograms using the Riesz transform with application to microscopy},
volume = {29},
month = {Oct},
year = {2012},
url = {https://opg.optica.org/josaa/abstract.cfm?URI=josaa-29-10-2118},
doi = {10.1364/JOSAA.29.002118},
}

@article{Sierra-Vazquez2010,
author = {Vicente Sierra-V\'{a}zquez and Ignacio Serrano-Pedraza},
journal = {J. Opt. Soc. Am. A},
number = {4},
pages = {781--796},
publisher = {Optica Publishing Group},
title = {Application of Riesz transforms to the isotropic AM-PM decomposition of geometrical-optical illusion images},
volume = {27},
month = {Apr},
year = {2010},
url = {https://opg.optica.org/josaa/abstract.cfm?URI=josaa-27-4-781},
doi = {10.1364/JOSAA.27.000781},
}

@article{Felsberg2002,
  title={The monogenic signal},
  author={Felsberg, Michael and Sommer, Gerald},
  journal={IEEE transactions on signal processing},
  volume={49},
  number={12},
  pages={3136--3144},
  year={2002},
  publisher={IEEE}
}

@article{Yamaguchi1997,
author = {Ichirou Yamaguchi and Tong Zhang},
journal = {Opt. Lett.},
number = {16},
pages = {1268--1270},
publisher = {Optica Publishing Group},
title = {Phase-shifting digital holography},
volume = {22},
month = {Aug},
year = {1997},
url = {https://opg.optica.org/ol/abstract.cfm?URI=ol-22-16-1268},
doi = {10.1364/OL.22.001268},
}

@article{Zhou2016,
  author={Zhou, Wenjing and Zhang, Hongbo and Yu, Yingjie and Poon, Ting-Chung},
  journal={IEEE Transactions on Industrial Informatics}, 
  title={Experiments on a Simple Setup for Two-Step Quadrature Phase-Shifting Holography}, 
  year={2016},
  volume={12},
  number={4},
  pages={1564-1570},
  doi={10.1109/TII.2015.2504843}}

@article{Zhou2009,
title = {Phase-shifting in-line digital holography on a digital micro-mirror device},
journal = {Optics and Lasers in Engineering},
volume = {47},
number = {9},
pages = {896-901},
year = {2009},
issn = {0143-8166},
doi = {https://doi.org/10.1016/j.optlaseng.2009.02.008},
url = {https://www.sciencedirect.com/science/article/pii/S0143816609000438},
author = {Wenjing Zhou and Qiangsheng Xu and Yingjie Yu and Anand Asundi},
}

@article{Ahrens2005,
  title={Paraview: An end-user tool for large data visualization},
  author={Ahrens, James and Geveci, Berk and Law, Charles},
  journal={The visualization handbook},
  volume={717},
  number={8},
  year={2005}
}

@book{Ayachit2015,
  title={The paraview guide: a parallel visualization application},
  author={Ayachit, Utkarsh},
  year={2015},
  publisher={Kitware, Inc.}
}

@article{2020SciPy-NMeth,
  author  = {Virtanen, Pauli and Gommers, Ralf and Oliphant, Travis E. and
            Haberland, Matt and Reddy, Tyler and Cournapeau, David and
            Burovski, Evgeni and Peterson, Pearu and Weckesser, Warren and
            Bright, Jonathan and {van der Walt}, St{\'e}fan J. and
            Brett, Matthew and Wilson, Joshua and Millman, K. Jarrod and
            Mayorov, Nikolay and Nelson, Andrew R. J. and Jones, Eric and
            Kern, Robert and Larson, Eric and Carey, C J and
            Polat, {\.I}lhan and Feng, Yu and Moore, Eric W. and
            {VanderPlas}, Jake and Laxalde, Denis and Perktold, Josef and
            Cimrman, Robert and Henriksen, Ian and Quintero, E. A. and
            Harris, Charles R. and Archibald, Anne M. and
            Ribeiro, Ant{\^o}nio H. and Pedregosa, Fabian and
            {van Mulbregt}, Paul and {SciPy 1.0 Contributors}},
  title   = {{{SciPy} 1.0: Fundamental Algorithms for Scientific
            Computing in Python}},
  journal = {Nature Methods},
  year    = {2020},
  volume  = {17},
  pages   = {261--272},
  adsurl  = {https://rdcu.be/b08Wh},
  doi     = {10.1038/s41592-019-0686-2},
}

\end{document}